%% file: Top3bw.tex
\documentclass[preprint,10pt]{elsarticle}

\usepackage{amssymb}
\usepackage{amsmath}
\usepackage{subcaption}
\RequirePackage{amssymb,amsfonts,amsthm}
\RequirePackage{mathrsfs}
\usepackage[left=30mm, right=45mm]{geometry}
\usepackage{hyperref}
\hypersetup{
colorlinks=true, %set true if you want colored links
linktoc=all,     %set to all if you want both sections and subsections linked
linkcolor=blue,  %choose some color if you want links to stand out
}

\newcommand {\Br}{\mathsf{Br}}
\newcommand \Lgr {\mathscr{L}}  % calligraphic
\newcommand {\calO}{\mathcal {O}}
\newcommand \mfrac[2]   {\displaystyle \frac{ #1}{#2} }  
\newcommand \decbw {t \to b \, \bar{b} \, b \, W^+}
\newcommand \hbpm {H^{\pm}}
\def  \gsim   {\raisebox{-3pt}{$\>\stackrel{>}{\scriptstyle\sim}\>$}}
\def  \lsim   {\raisebox{-3pt}{$\>\stackrel{<}{\scriptstyle\sim}\>$}} 

\begin{document}

\begin{frontmatter}



% pdflatex HiggsCouplings.tex; bibtex HiggsCouplings; pdflatex HiggsCouplings.tex;

  \title{Charged Higgs contribution to rare top-quark decay
    $t \to b \bar{b} b W^+$ }


\author[label1]{S.~Slabospitskii}

\address[label1]{NRC ``Kurchatov Institute'' - IHEP, Protvino, Moscow Region, Russia}
 
%\address{}
\ead{Sergei.Slabospitskii@ihep.ru}

\begin{abstract}
  The calculation of the rare $t$-quark decay $\decbw$
  within the Standard Model as well as the charged Higgs contribution
  to this decay is presented.
  The role of possible background processes is discussed.
  It is shown that this decay provides an additional
  possibility for charged Higgs searches. 
  \\ [5mm]
\end{abstract}


\end{frontmatter}


\input{top_decay}

\bibliographystyle{elsarticle-num} 
\biboptions{numbers,sort&compress}
\bibliography{top_biblio.bib}

\end{document}

%% file: top_decay.tex
\section{Introduction}
\label{sec:intro}

The physics of the top-quark is one the main areas  of study
in current high energy physics. Indeed almost all characteristics
of the processes with
top-quark can be calculated with high theoretical accuracy.
The  decay $t \to b W$ is by far the dominant one with the Standard Model
(SM). All other decay channels have very small branching
fractions and are predicted to be smaller by
several orders of magnitude in the SM~\cite{Beneke:2000hk, Zyla:2020zbs}.

 On the other hand, rare decays of the $t$-quark
are very sensitive to the manifestation of New Physics beyond
the SM. A very well known example is top-quark decays due to 
anomalous flavor changing neutral
currents~\cite{Beneke:2000hk}.

In this article we study a rare top-quark decay: 
\begin{eqnarray*}
  t \, \to \, b \, W^+ \, \bar{b} \, b 
\end{eqnarray*}
as well as  the charged Higgs boson
contribution to this decay.
It is shown that this decay  has well identified
final states  and has the 
relatively large branching
fraction, $\Br(\decbw) \sim \calO(10^{-3})$.  
 We demonstrate that this four-body decay channel
provides an additional way to search for charged $\hbpm$-boson.

Charged Higgs bosons appear in the scalar sector of several
SM extensions, and is
the object of various beyond the Standard Model %(BSM)
searches at the LHC. In this article we explore
a generic two-Higgs-doublet model (2HDM),
which is one of the simplest SM
extensions featuring a charged scalar~\cite{Barger:1987nn, Branco:2011iw}.
Within this class of models,
two isospin doublets are introduced to
break the $SU(2) \times U(1)$ symmetry, leading to the existence of
five physical Higgs bosons, two of which
are charged particles ($H^{\pm}$).
The latest constraints the allowed $H^{\pm}$ mass range as a function of
$\tan \beta$ can be found in~\cite{Akeroyd:2016ymd, Arbey:2017gmh}.

Searches for $H^{\pm}$have been performed at
LEP~\cite{ALEPH:2013htx}, at the Fermilab
Tevatron~\cite{CDF:2009esh,D0:2011rhz}. 
The ATLAS and CMS Collaborations have covered several
$H^{\pm}$ decay channels, such as
$\tau \nu_{\tau}, \;  t b, \;  c s, \;  c b$
(see~\cite{ATLAS:2018fwl, ATLAS:2018gfm, ATLAS:2021upq, CMS:2019eih,
  CMS:2020osd, CMS:2019rlz, CMS:2020imj} and references therein).
Note, that  $\hbpm$-boson (like a SM Higgs) has couplings proportional to
fermion masses.

The large value of the $t \, \hbpm \, b$ coupling leads
to the fact that processes involving top quarks are considered
for the charged Higgs production processes.
The  experiments explore three scenarios for the
search for a charged Higgs boson.
For ``light'' $\hbpm$-boson ($M(\hbpm) < m_t$) top-quark decay channel
into charged Higgs is used.
For the case when $M(\hbpm)$ is greater than $t$-quark mass, it is
considered the charged Higgs production with 
subsequent decay into $t \bar{b}$ state.
As a result the charged Higgs with masses
\begin{eqnarray}
  M(\hbpm) < 160 \;\; \hbox{GeV} \quad \hbox{and} \quad
  M(\hbpm) > 180 \;\; \hbox{GeV}
  \label{eq-exp-1}
  \end{eqnarray}
with a wide range of $\tan \beta$ are
excluded~\cite{ATLAS:2018fwl, ATLAS:2018gfm, ATLAS:2021upq, CMS:2019eih,
  CMS:2020osd, CMS:2019rlz, CMS:2020imj}.

The third scenario explores the production of the Higgs boson
in association with a top quark (see~\cite{ATLAS:2018gfm}) with 
subsequent $\hbpm \to \tau^{\pm} \nu_{\tau}$ decay. As a result,
the charged Higgs with mass range $(90-2000)$~GeV
and $\tan \beta \gsim 1$ is excluded.

Note, that in all experiments, a search is made for the direct production
of charged Higgs with subsequent decays in observed fermions.
In the present articles we propose an additional
``indirect'' way for $H^{\pm}$ searches by study
of a  rare four-body top-quark decay channel
($t \to b W^+ b \bar{b}$). It is shown, that 
$t$-quark decay~($t \to bW^+b\bar{b}$) provides a reasonable
possibility to detect charged Higgs contribution.

Bearing in mind the experimental constraint~(\ref{eq-exp-1})
in the present article
we investigate the role of the charged $\hbpm$-boson 
with mass range and $\tan \beta$ as follows:
\begin{eqnarray}
   M(\hbpm) = (160 \, \div 180) \;\; \hbox{GeV} \quad \hbox{and} \quad
 \tan \beta \lsim 1
  \label{eq-hrange}
\end{eqnarray}

Throughout of this  article we follow~\cite{Borodulin:2017pwh}
for the notations, the  SM vertices and SM parameters.
For numerical calculations of the  decay widths and 
production processes the C++ version of the {\sf TopReX}
package~\cite{Slabospitsky:2002ag} is used. 
All calculations were done the following  quark masses: 
\begin{eqnarray}
  \left.
  \begin{array}{lrlr}
  m_d = & 0.33 \;\; \hbox{GeV}, &   
  m_u = & 0.33 \;\; \hbox{GeV},
  \\
  m_s = & 0.5 \;\; \hbox{GeV}, &   
  m_c = & 1.5 \;\; \hbox{GeV},
  \\
  m_b = & 4.8 \;\; \hbox{GeV}, &   
  m_t = & 172.5 \;\; \hbox{GeV},
  \end{array}
  \right.
  \label{eq-1}
\end{eqnarray}

We explore the $b$-jet tagging with $b$-quark identification efficiency of
70\%, with probability to misidentify $c$ quark, and light-flavor
quark and gluon jets as $b$-jets of approximately 10\% and 1.\%,
respectively  (see, e.g~\cite{CMS:2017wtu}):
\begin{eqnarray}
\epsilon_b = 0.7, \epsilon_c = 0.1,
\epsilon_{q, \; g} = 0.01
 \label{eq-btag}
\end{eqnarray}

The article is organized as follows.  Charged Higgs
interaction model is described in the Section~\ref{lagrangian}.
The constraints on $\tan \beta$ evaluated with taking into account
for $t$-quark total decay width, as well as the branching
fractions of $t$-quark decays, is presented in the
Section~\ref{constraints}.
The width of the top-quark decay $t \to b \, \bar{b} \, b \, W $
is calculated in the Section~\ref{decay3bw}.
A brief comment on $t$-quark decay
to a multilepton final state is given in the Section~\ref{fourleptons}.
The different background sources are considered in the
Sections~\ref{background1} and~\ref{background2}.
The kinematic of the decay
$t \, \to \, b \, \bar{b} \, b \, W^+$ is discussed in the
Section~\ref{kinematics}.
The last Section~\ref{conclusion} summarize the obtained results.

%%%%%%%%%%%%%%%%%%%%%%%%%%%%%%%%%%%%%%%%%%
\section {Charged Higgs interaction Lagrangian  }
\label{lagrangian}

The interaction Lagrangian describing the
$H^{\pm}$-boson-fermions interactions  in the MSSM
is (see \cite{Beneke:2000hk}, \cite{Branco:2011iw}, \cite{Barger:1987nn}): 
    
\begin{eqnarray}
{\cal L} & = &\frac{g}{\sqrt{2}M_{W}} H^+
 \{V_{ud} \overline u (m_u \cot \beta P_L + m_d \tan\beta P_R) d +  \, 
 \overline{\nu} (\tan\beta m_{\ell} P_R) \ell\},
 \label{eq-2}
\end{eqnarray}
where $P_{L/R} = \frac{1}{2} \left( 1 \mp \gamma^5 \right)$,
symbols $u$ and $d$ stand for "up" ($u, c, t$) and "down" ($d, s, b$)
quarks,  
$\nu$ and $\ell$ for neutrino and charged leptons,
 $V_{ud}$ 
is the Cabibbo-Kobayashi-Maskawa matrix element.

At tree level the corresponding decay widths are equal
to~\cite{Beneke:2000hk}:
\begin{eqnarray}
\Gamma\left(H^{+} \rightarrow l \nu \right) &=&
    \frac{g^{2}M_{H}}{32\pi M^{2}_{W}}m^{2}_{l}\tan^{2}\beta
    \label{eq-3}
 \\
\Gamma\left(H^{+} \rightarrow q\bar{q}\right) & = &
    \frac{3g^{2}}{32\pi M^{2}_{W}M_{H}}
    \left|V_{q\bar{q}}\right|^{2}\lambda^{1/2}
    \left(1,\frac{m^{2}_{q}}{M^{2}_{H}},
    \frac{m^{2}_{\bar{q}}}{M^{2}_{H}}\right)\times \nonumber
    \\
 &&   \left[
       \left(M^{2}_{H} - m^{2}_{\bar{q}} - m^{2}_{q}\right)
       \left(m^{2}_{q}\cot^{2}\beta + m^{2}_{\bar{q}}\tan^{2}\beta\right)
        - 4m^{2}_{q}m^{2}_{\bar{q}}
    \right]
 \label{eq-4}
 \\
\Gamma\left(t \rightarrow bH^{+}\right) & = &
    \frac{g^{2}}{64\pi M^{2}_{W}m_{t}}
    \left|V_{tb}\right|^{2}\lambda^{1/2}
    \left(1,\frac{m^{2}_{b}}{m^{2}_{t}},
    \frac{M^{2}_{H}}{m^{2}_{t}}\right)\times \nonumber
    \\
 &&   \left[
       \left(m^{2}_{t} + m^{2}_{b} - M^{2}_{H}\right)
       \left(m^{2}_{t}\cot^{2}\beta + m^{2}_{b}\tan^{2}\beta\right)
        + 4m^{2}_{t}m^{2}_{b}
    \right]
 \label{eq-5}
\end{eqnarray}
where 
\[ \lambda\left(a,b,c\right)=a^{2}+b^{2}+c^{2}-2\left(ab+ac+bc\right)
\] 

\begin{figure}[h!]
  \begin{center}
\includegraphics[width=0.28\textwidth,clip]{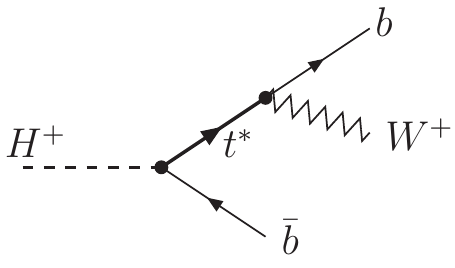} 
\end{center}
\vspace{-5mm}
\caption{Diagram describing $H^+ \to t^* \bar{b} \to W^+ b \bar{b}$
  decay due to virtual $t$-quark contribution.
  \label{fig:hbbw}
  }
\end{figure}

Note, that for small $\tan \beta$ the additional decay channel
($H^+ \to t^* \bar{b} \to W^+ b \bar{b}$) can get a
noticeable contribution to the total charged Higgs decay
width~\cite{Ma:1997up} (see diagram in Fig.~\ref{fig:hbbw}).

\begin{figure}[h!]
  \begin{center}
\includegraphics[width=0.48\textwidth,clip]{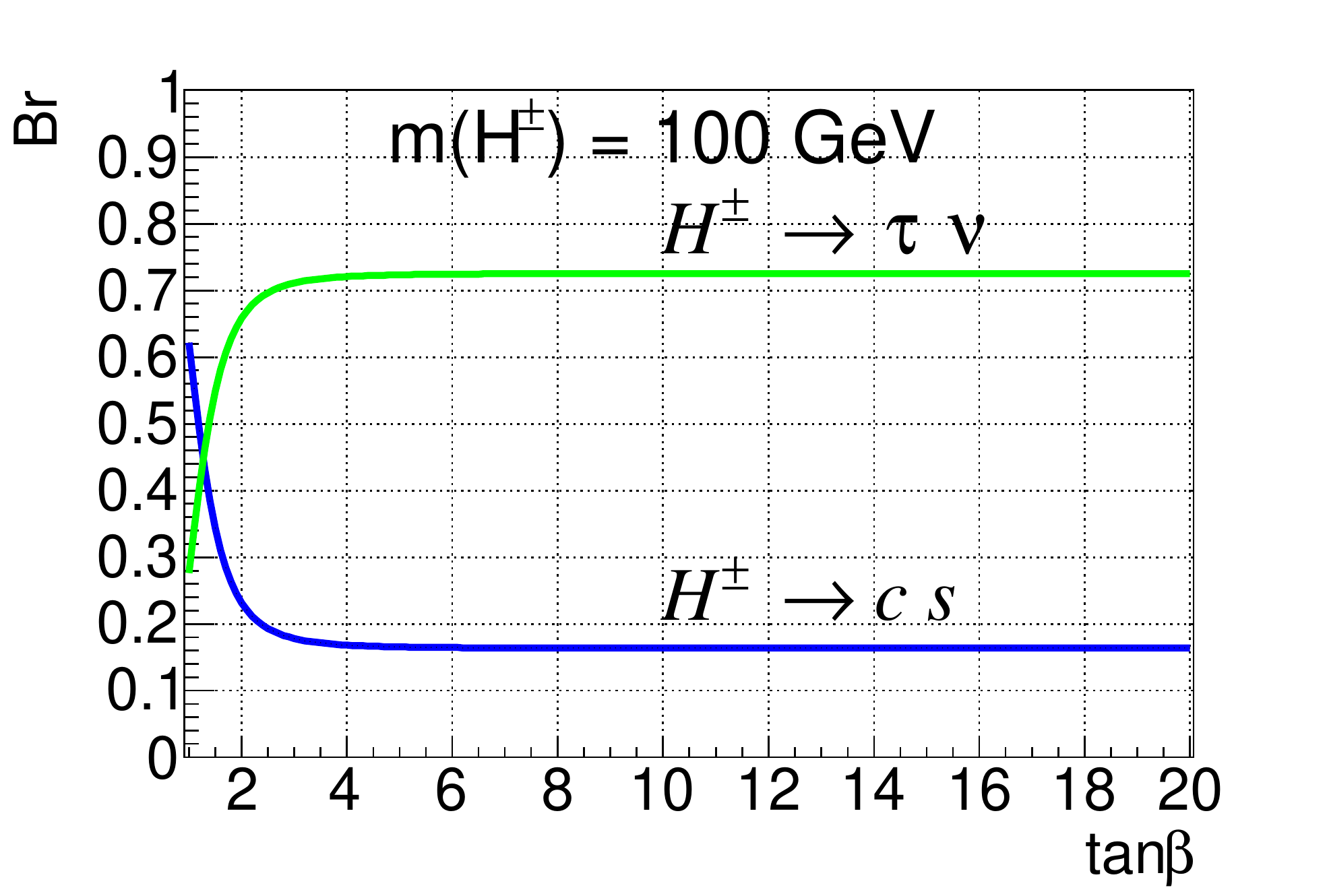} 
\includegraphics[width=0.48\textwidth,clip]{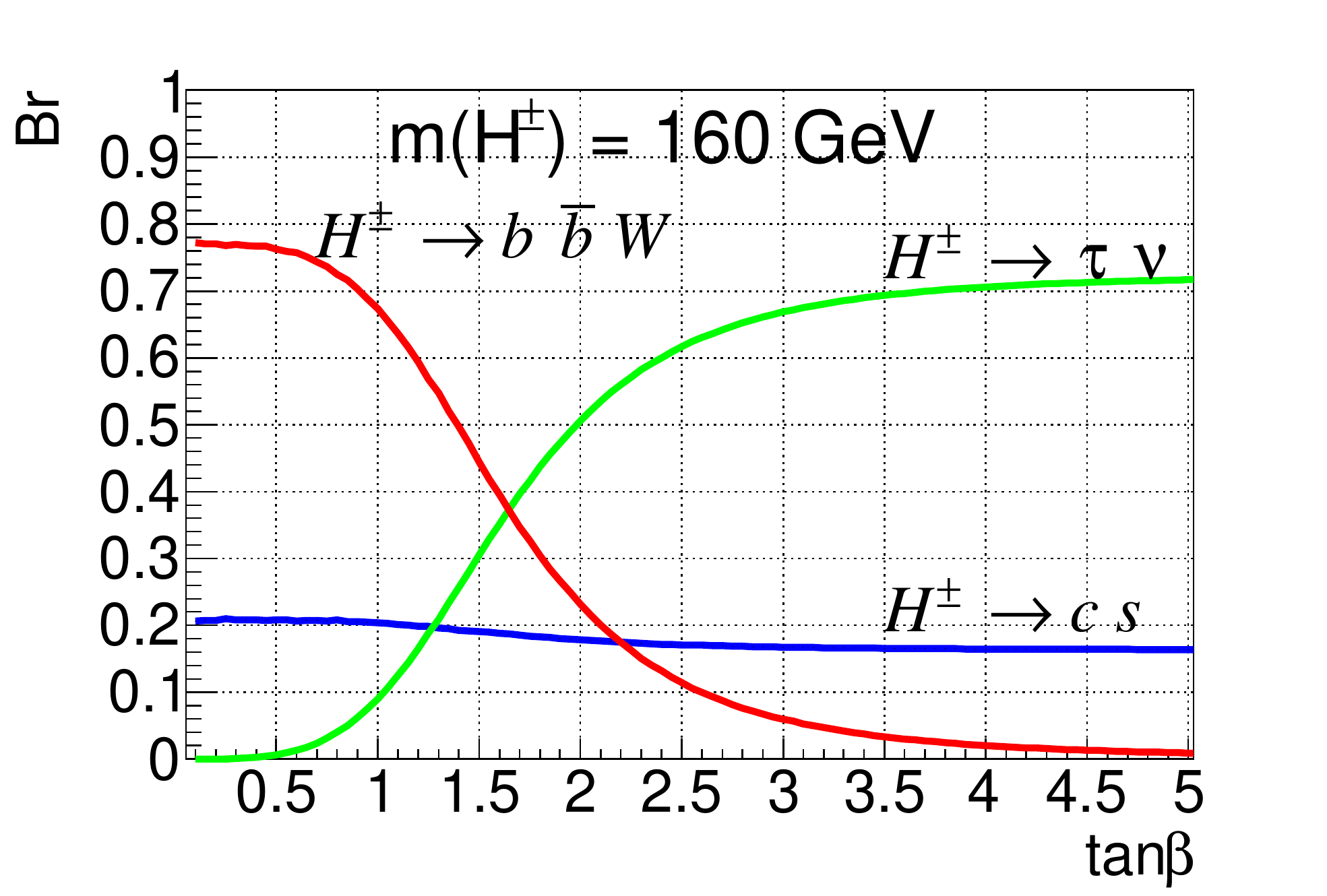} 
\\
\includegraphics[width=0.48\textwidth,clip]{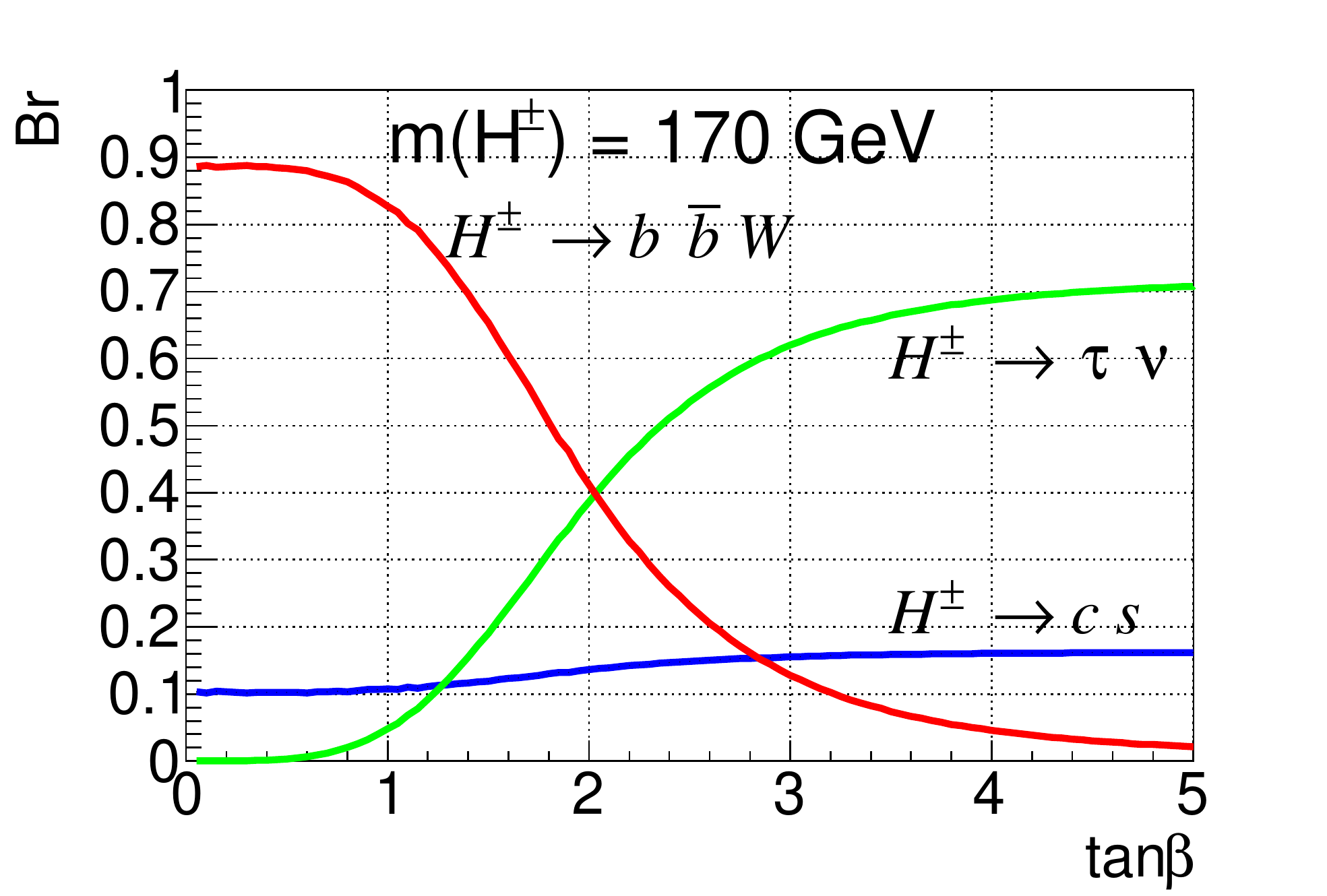} 
\includegraphics[width=0.48\textwidth,clip]{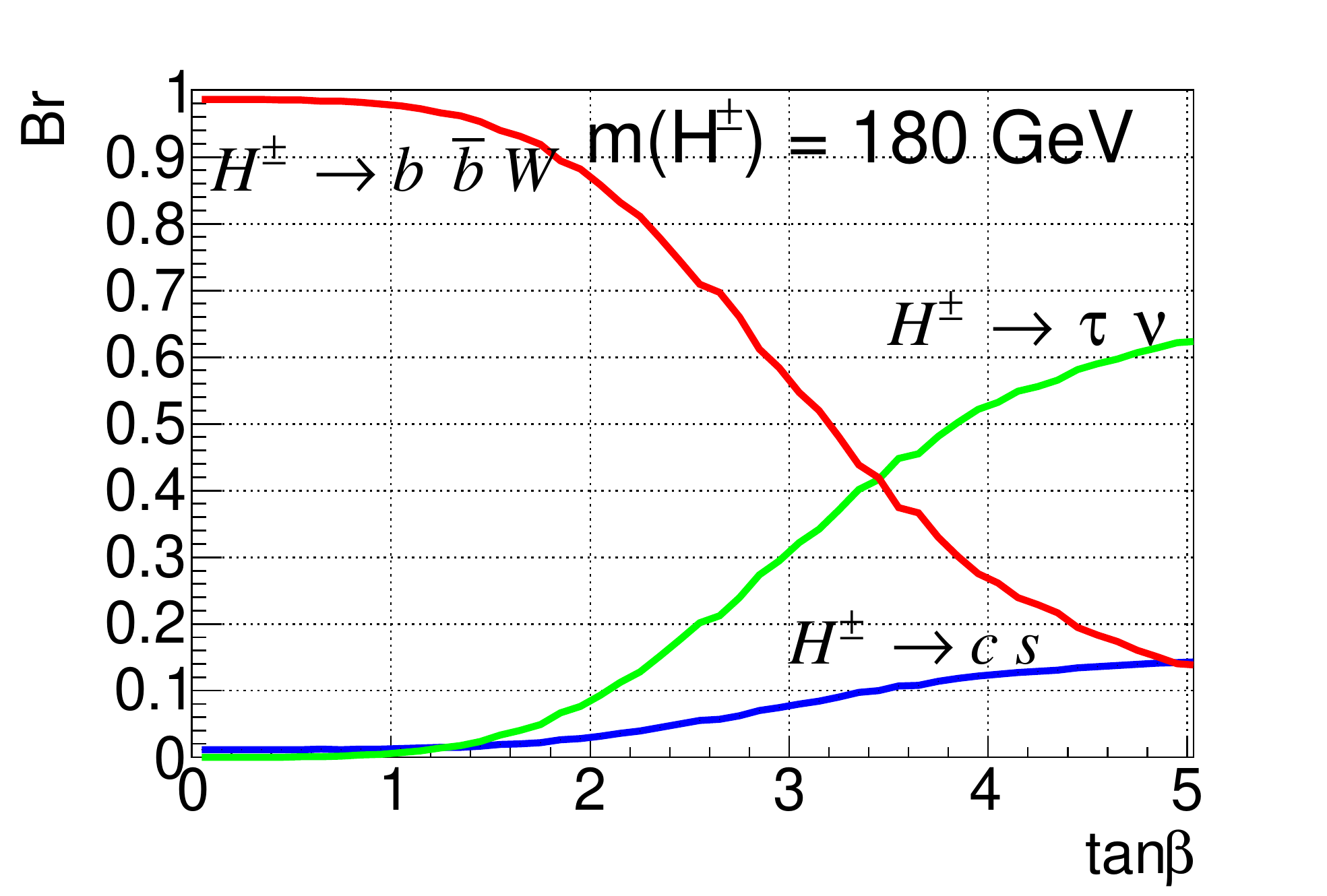} 
\end{center}
\vspace{-5mm}
\caption{Branching fractions for three $H^{\pm}$ decay modes
  for four  values of $m_{H^{+}}$ vs. $\tan \beta$.
  \label{fig:brhpm}
  }
\end{figure}

In the Fig.~\ref{fig:brhpm} we present the behavior of the
branching fractions for three $H^{\pm}$ decay modes
($c \bar{s}, \, \tau \nu_{tau}, \; b\bar{b}W$) 
  for four  values of $m_{H^{+}}$ versus $\tan \beta$.
  As it seen for small $m_{H^{+}}$ values ($m_{H^{+}} \sim 100$~GeV) this
  three-body decay ($H^+ \to b \bar{b} W^+$) is almost negligible.
Consequently, for these $m_{H^+}$-values  one  expects 
$H^+ \to \tau^+ \nu$ (favoured for large $\tan\beta$) and/or 
$H^+ \to c \bar{s}$ (favoured for small $\tan\beta$) to be the 
dominant decays. 

On the other hand, if $\tan \beta \lsim 2$ and $m_{H^{+}} \gsim 150$~GeV,
  the large mass (or coupling) of the $t$-quark causes
  $\Br(H^{+} \to b \bar{b} W^+)$ to exceed
  $\Br(H^{+} \to c \bar{s})$. Moreover, for $\tan \beta < 1$ this
  three-body decay mode become dominant~\cite{Beneke:2000hk, Ma:1997up} 
  (see Fig.~\ref{fig:brhpm}).
  
\section{Constraints on charged Higgs parameters}
\label{constraints}

As explained in the introduction, we consider the
charged Higgs boson with mass
from 160 GeV to 180~GeV and small values of $\tan\beta$
(this area has not yet been studied in detail in the experiments).

In this section we present very rough estimates for  the limits
of the allowed region   of $\tan \beta$ values.
The charged Higgs (if it exists) should make a contribution
into the total $t$-quark decay width as well as to the
branching fractions. 
These values are taken from 
``Review of Particle Properties''~\cite{Zyla:2020zbs}).

\begin{eqnarray}
  \Gamma_t & =&  1.42^{+0.19}_{-0.15} \;\;\; \hbox{GeV}
     \label{gamma_exp}
  \\
   \Br(t \to b q \bar{q}) & = & (66.5 \pm 1.4) \%
 \label{br-bqq}
  \\
  \Br(t \to b \, \tau \nu_{\tau}) & =& (11.1 \pm 0.9) \%
  \label{br-btau}
\end{eqnarray}

The first constraints could be obtained from requirement that
the contribution to the total
top-quark decay width could not
exceed $\Delta \Gamma_t^{H} \approx 0.2$~GeV.
Additional constrains could be deduced be using of the
top-quark branching fractions (\ref{br-bqq}) and (\ref{br-btau}).

Using the errors of these branching fractions we require that
charged Higgs contribution to branching fractions of these
decays ($t \to b q \bar{q}$ and
$t \to b \, \tau \nu_{\tau}$) should be less then the errors
from~(\ref{br-bqq}-\ref{br-btau}).
Therefore, we explore the constraints as follows:
\begin{eqnarray}
 (1) && \Delta \Gamma_t^{H} < 0.2 \;\; \hbox{GeV}
  \label{constraint-1}
  \\
(2) &&  \left.
  \begin{array}{lcl}
    \Delta \Br^{(H)}(t \to b q \bar{q}) =
    \mfrac {\Gamma^{(H)}(t \to b q \bar{q})}
          {\Gamma_{LO}+ \Gamma^{(H)}(t \to b H^{+})}
          & < & 0.014
          \\
    \Delta \Br^{(H)}(t \to b \tau \nu_{tau}) =
    \mfrac {\Gamma^{(H)}(t \to b q \bar{q})}
          {\Gamma_{LO}+ \Gamma^{(H)}(t \to b  H^{+})}
          & < & 0.009
  \end{array}
  \right\}
  \label{constraint-2}
\end{eqnarray}
where $\Gamma_{LO}$ is  $t$-quark decay width evaluated at the leading
order:
$\Gamma_{LO} = \Gamma( t \to b \, W^+) = 1.47$~GeV~\cite{Beneke:2000hk}

\begin{figure}[h!]
  \begin{center}
\includegraphics[width=0.7\textwidth,clip]{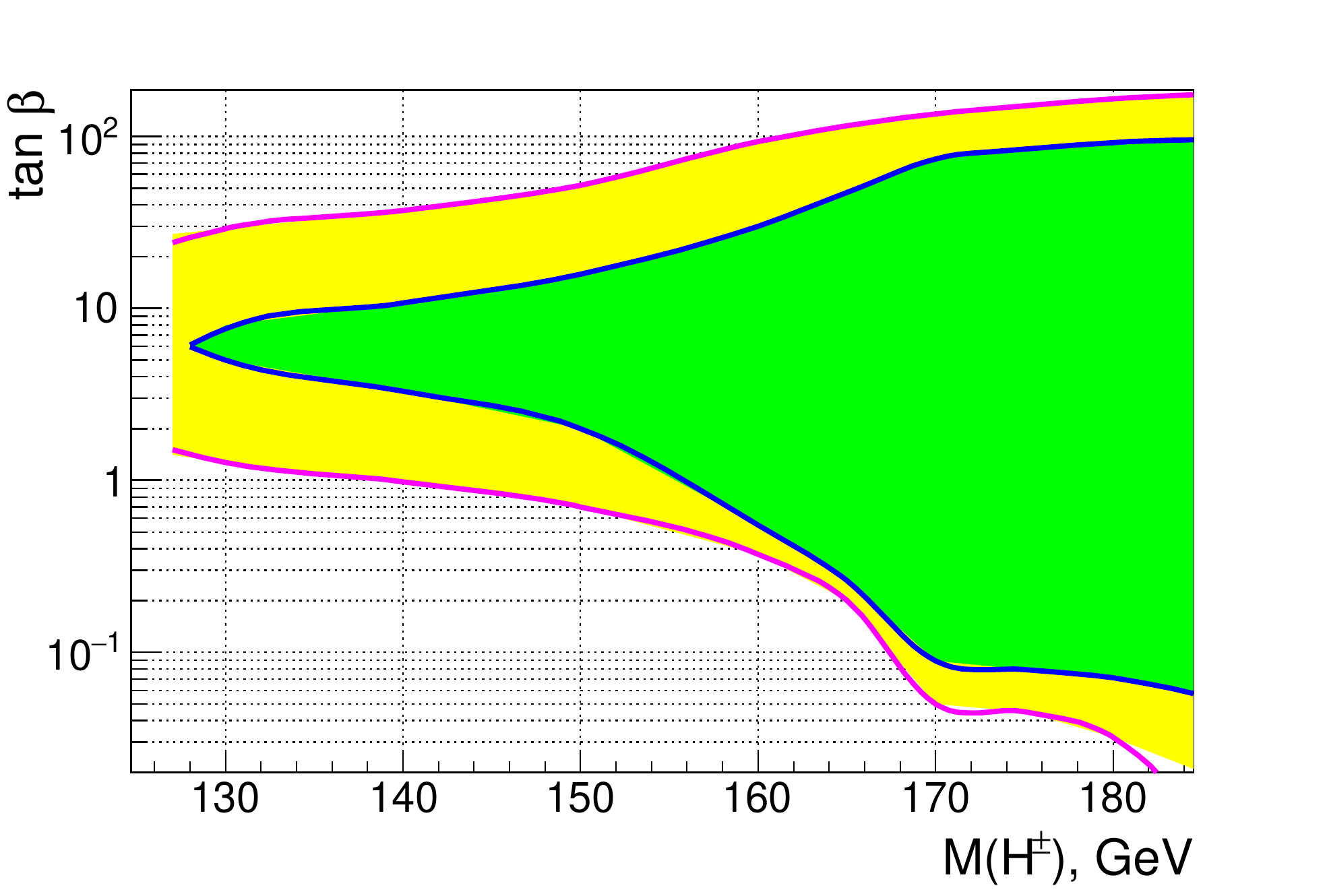} 
 \end{center}
\vspace{-5mm}
\caption{The allowed charged Higgs parameters range. The wide
  area correspond to the constraint (\ref{constraint-1}),
  while the small one is resulted by using
  of the constraints~(\ref{constraint-2}).
   \label{fig:limits}
  }
\end{figure}

   The resulted allowed parameter ranges are presented in the
  Table~\ref{tab-1}
  and in Fig.~\ref{fig:limits}.
%  This analysis was done for  mass range
%  ($M(\hbpm) \gsim 120$~GeV).
 It follows form this table and figure that the more  ``narrow''
 limits are resulted from the branching fraction  constraints.
 
\begin{table}[h!] 
  \caption{
    The allowed charged Higgs parameters range. The symbols (1) and (2)
    corresponds to the constraints (\ref{constraint-1})
    and (\ref{constraint-2}). The mass of $H^{\pm}$-boson is in GeV.
     }
  \label{tab-1}
\begin{center}
 \renewcommand{\arraystretch}{1.1}
 \begin{tabular}{c||c|c||c||c|c}
 $M(H^{\pm})$ & \multicolumn{2}{c||}{ $\tan \beta$ } &
 $M(H^{\pm})$ & \multicolumn{2}{c}{ $\tan \beta$ }
  \\ \hline 
      & (1)  &  (2) &  & (1)  & (2)  
  \\ \hline
  $120$   & $1.50 \div 24.$ & - &
  $160$  & $0.37 \div 93$   & $0.55  \div 30.$
   \\ \hline
   $125$   & $1.4 \div 27.$ & $5.9 \div 6.1$ &
   $165$  & $0.20 \div 115$  & $0.26  \div 47$
    \\ \hline
  $130$   & $1.27 \div 29.$ & $5. \div 8.$ &
$170$  & $0.05 \div 135$  & $0.089  \div 74$
\\ \hline
$140$  & $0.98 \div 37$   & $3.3  \div 10.7$
  & $175$  & $0.045 \div 150$ & $0.079  \div 84$
\\ \hline
$150$  & $0.70 \div 52$   & $2.0  \div 15.8$ &
$180$  & $0.032 \div 165$ & $0.071  \div 92$
     \end{tabular}
\end{center}
\end{table}
\renewcommand{\arraystretch}{1.}

Certainly, these results  should be considered as  very rough
estimates. To get more correct results one needs to
take into account the NLO calculations and the running masses of the quarks.

Moreover, we do not compare these limits (see~Table~\ref{tab-1}) with
results from~\cite{Akeroyd:2016ymd, Arbey:2017gmh},
because, in fact, we use a rather narrow "working" area of
$\tan \beta$ parameter 
(see Section~\ref{decay3bw} and Fig.~\ref{fig:limits_dec}).
We use these range for $M(\hbpm)$~(\ref{eq-hrange})
and  $\tan \beta$  in  further calculations.
In addition, for illustration, we present the total decay widths of the
charged Higgs boson $\Gamma_{tot}(\hbpm)$ calculated within these
limits (see Table~\ref{tab-2}).

\begin{table}[t!] 
  \caption{
  The total charged Higgs boson decay width  $\Gamma_{tot}(H^{\pm})$
}
  \label{tab-2}
  
\begin{center}
 \renewcommand{\arraystretch}{1.2}
 \begin{tabular}{l||c|c|c|c|c}
  $ \tan \beta$ & $160$ & $165$  & $170$ & $175$ & $180$ 
     \\ \hline
 $ 0.1$ & - &  -  &   $0.62$  &  $1.18$ & $6.96$ \\ \hline
$ 0.2$ & - & -  &   $0.154 $  &  $0.30$ & $1.75$  \\ \hline
$0.35$ & -  & $0.031$ &   $0.051 $  &  $0.096$ & $0.051$ \\ \hline
$ 0.5$ & $0.010$ & $0.0154$ & $0.025 $  &  $0.048$ & $0.028$  
     \\ \hline
 $ 1$ & $0.0028$ &$0.004 $  &  $0.0065$ & $0.0132$ & $0.022$
     \\ \hline
 $ 10$ & $0.046$ & $0.047 $  &  $0.049$ & $0.051$ & $0.058$ 
     \\ \hline
$ 50$ & - & $1.18$  &  $1.22$ & $1.26$ & $1.51$ 
       \end{tabular}
\end{center}
\end{table}
\renewcommand{\arraystretch}{1.}

%%%%%%%%%%%%%%%%%%%%%%%%%%%%%%%%%%%%
\section{Top-quark decay $t \to b \, W^+ \, b\,  \bar{b} $}
\label{decay3bw}
 
\begin{figure}[h!]
 \begin{center}
   \includegraphics[width=0.75\textwidth,clip]{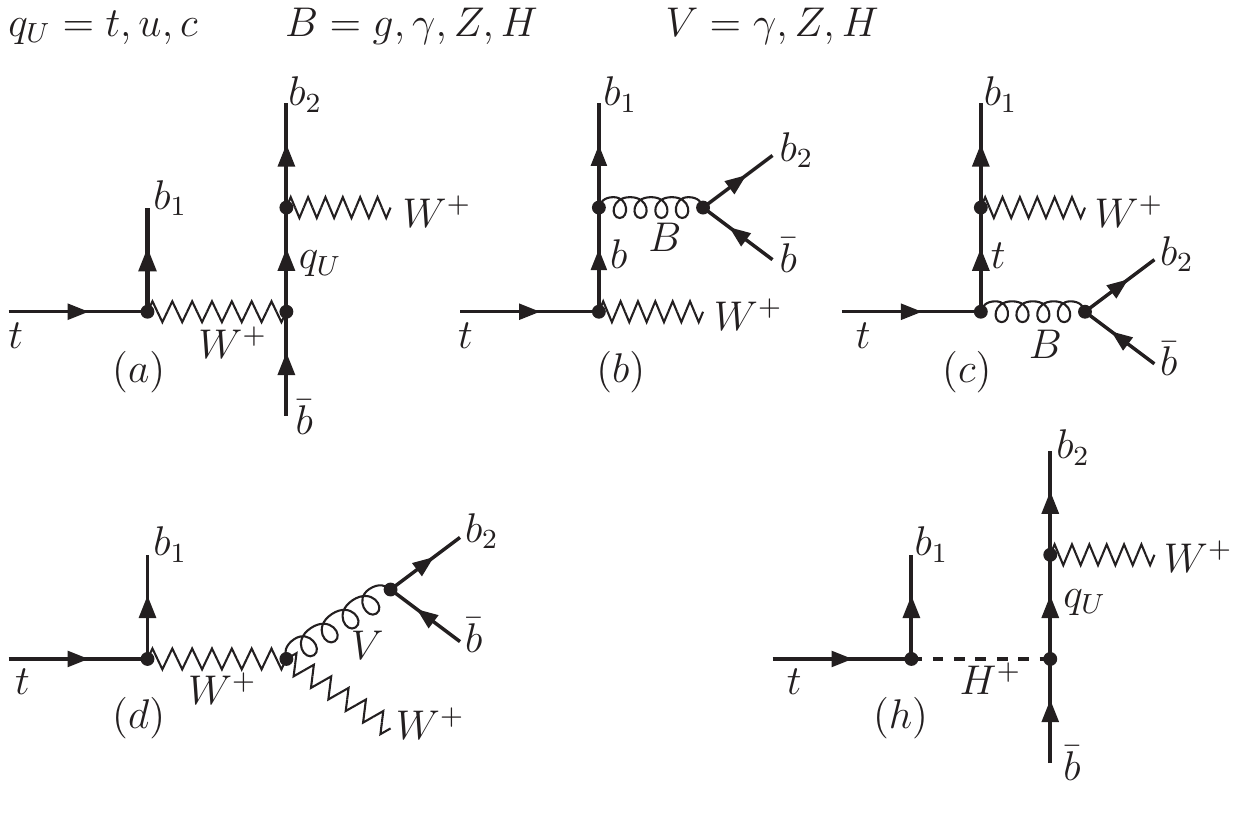}
\end{center}
\vspace{-5mm}
\caption{The diagrams($(a), (b), (c)$ and $(d)$) describing
  $t \to b W^+ \; b \bar{b}$ decay within SM.
  The diagram $(h)$ describes charged Higgs contribution to this decay.
   \label{fig:diagrams}
  }
\end{figure}

In this article we consider charged Higgs contribution to rare four-body
top-quark decay:
\begin{eqnarray}
t \to b W^+ \; b \bar{b}
 \label{eq-dec3bw}
\end{eqnarray}
Within SM this decay process is described by 28 Feynman diagrams
(see Fig.~\ref{fig:diagrams}). The picture~$(a)$ corresponds to
three diagrams with virtual quarks $q_U$ exchange ($q_U = t, \; u$
or $q_U = c$).

In the same fashion the pictures~$(b)$ and $(c)$ correspond to
eight diagrams with virtual bosons $B$ exchange ($B = g$, $B = \gamma$,
$B = Z$ or $B = H$).
 The picture~$(d)$ corresponds to
three diagrams with virtual bosons $V$ exchange ($V = \gamma$, $V = Z$
or $V = H$).
 The picture $(h)$ corresponds to
 three diagrams with charged Higgs and with virtual
 quarks  $q_U$ exchange ($q_U = t, \, u, \, c$).
Note, that the light virtual quarks ($u$ and $c$)
as well  virtual photon, $Z$ and $H$ bosons give
very small contributions to the
$t \to b W^+ \; b \bar{b}$ decay width.

 The calculated decay width and branching fraction are as follows: 
\begin{eqnarray}
  \left.
  \begin{array}{lcl}
  \Gamma(t \to b W^+ b \bar{b}) &=& ( 9.30 \pm 0.03) \times 10^{-4}
  \;\;\; \hbox{GeV}
  \label{eq-8}
  \\
  \Br(t \to b W^+ b \bar{b}) & =& (6.29 \pm 0.02) \times 10^{-4}
  \end{array}
  \right.
  \label{eq-9}
\end{eqnarray}

%%%%%%%%%%%%%%%

In the  Table~\ref{tab-3} we present the results   for
several charged Higgs mass values. As it follows from calculations 
 the partial
decay width for three variants of the parameters
($\huge\{M(H^+) = 160$~GeV,
$\tan \beta < 0.35\huge\}$, $\huge\{M(H^+) = 165$~GeV,
$\tan \beta < 0.2\huge\}$,
and $\huge\{M(H^+) = 170$~GeV,
$\tan \beta<  0.05\huge\}$) is exceed value 0.2~GeV. Consequently,
this parameter space range should also be excluded
from our further discussion.

\begin{figure}[h!]
  \begin{center}
\includegraphics[width=0.7\textwidth,clip]{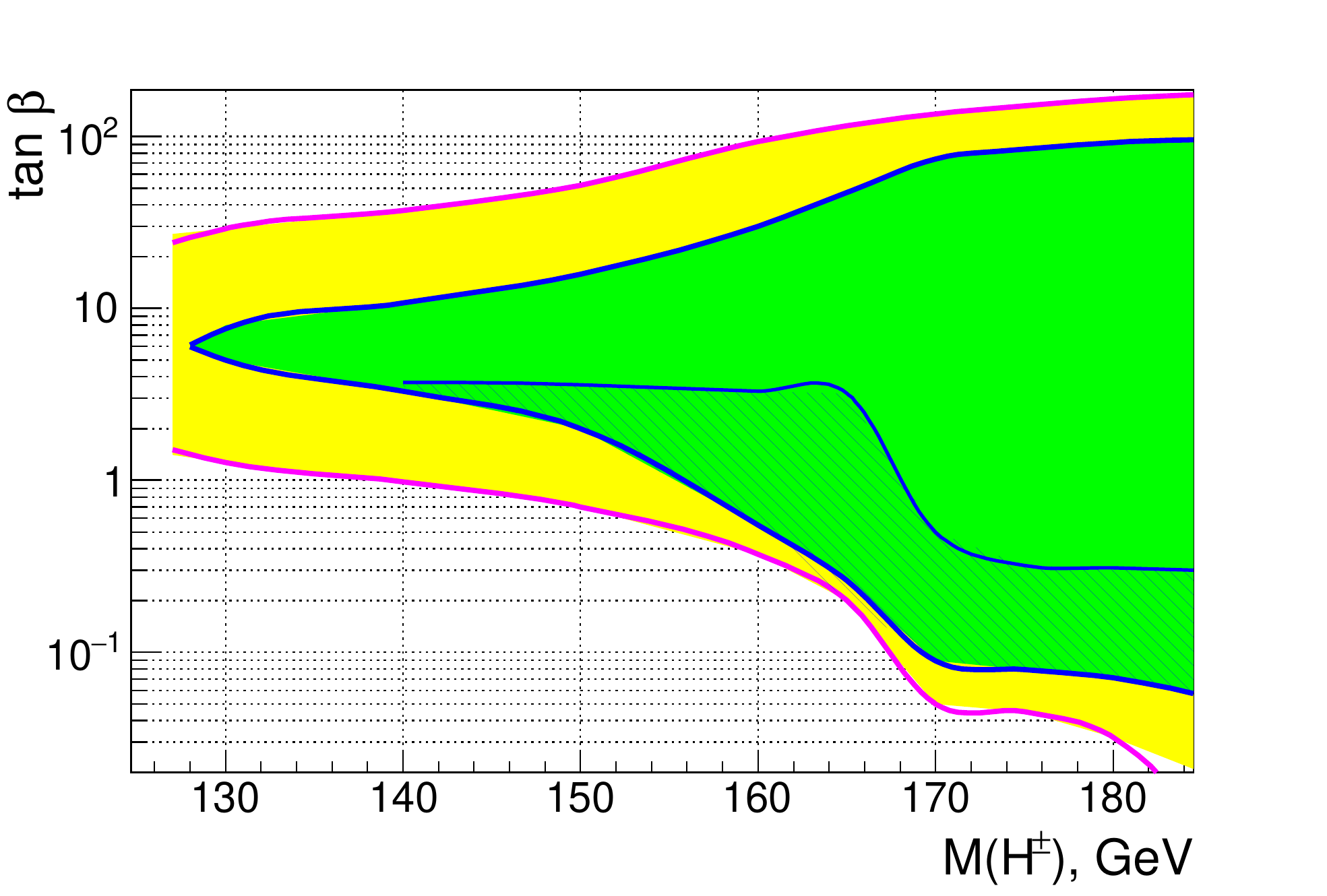} 
 \end{center}
\vspace{-5mm}
\caption{The same plot as in Fig.~\ref{fig:limits}
  (the allowed charged Higgs parameters range).
 The hatched  area corresponds to cases
    when the charged Higgs contribution to the $\decbw$ decay width
    exceeds 5\%.
     \label{fig:limits_dec}
  }
\end{figure}

\begin{table}[h!] 
  \caption{
    The partial decay widths of the top-quark decay $t \to b \bar{b} b W$
    with account of $H^{\pm}$ contribution.
    All width values (in GeV) are multiplied by $10^4$.
    The ``SM'' means that charged Higgs contribution
    is less the 5\% and the corresponding width equals
    $\Gamma_{SM} \approx 9.3 \times 10^{-4}$ GeV. For
    $tan \beta > 50$ and $M_H \ge 170$~GeV the decay widths are
  equal to $\Gamma_{SM}$}
    \label{tab-3}
  
\begin{center}
 \renewcommand{\arraystretch}{1.2}
{

 \begin{tabular}{l||c|c|c|c|c|c}
$\tan \beta$  &  $160$ & $165$  & $170$ & $175$ & $180$ & $185$ 
   \\ \hline
   $ 0.1$  & -  & - & $282 \pm 2.4 $ & $122 \pm 1.2$ & $75.1 \pm 0.6$
   & $50.6 \pm 0.4$
    \\ \hline 
 $ 0.2$    & - & -  & $28.6 \pm 0.0.2$ & $17.7 \pm 0.23$ &
    $ 14.5 \pm 0.20$    & $13.1 \pm 0.20$  
    \\ \hline 
   $ 0.25$ & - & $1631 \pm 211$ & $17.8 \pm 0.13$ & $13.1 \pm 0.1$ &
    $ 11.7 \pm 0.09$    & $11.0 \pm 0.09$  
    \\ \hline 
$ 0.5$ & $1110\pm 340$ &$388 \pm 47$  & $10.3 \pm 0.19 $
    & {\sf SM}  &  {\sf SM} & {\sf SM} %$9.38 \pm 0.18$  
    \\ \hline 
    $1$ & $254 \pm 84$& $114 \pm 37$ & {\sf SM} & {\sf SM}  &  {\sf SM}
    & {\sf SM} 
       \\ \hline 
 $30$ & $9.9 \pm 0.09$ & {\sf SM}   & {\sf SM}  & {\sf SM}  &
       {\sf SM}  & {\sf SM}
       \\ \hline 
 $50$ & -     & $10.3 \pm 0.11$  & {\sf SM}  & {\sf SM}  &
       {\sf SM}  & {\sf SM}
  \end{tabular}
 }
\end{center}
\end{table}
\renewcommand{\arraystretch}{1.}
The Fig.~\ref{fig:limits_dec} presents the same plot as
in Fig.~\ref{fig:limits}
  (the allowed charged Higgs parameters range).
 The hatched  area corresponds to cases
    when the charged Higgs contribution to the $\decbw$ decay width
    exceeds 5\%.

There is one important comment concerning the final $W$-boson decay
into observed particles. 
For $W$-boson decaying into  lepton pair ($W \to \ell \nu$)
 a sharp peak near the $W$-boson should be expected, and the corresponding
 decay width is
\begin{eqnarray*}
\Gamma( t \to b \bar{b} b \, \ell \nu) = 
\Gamma( t \to b \bar{b} b \, W^+) \times
\Br(W^+ \to \ell^+ \nu)
\end{eqnarray*}

\begin{figure}[h!]
 \begin{center}
  \includegraphics[width=0.75\textwidth,clip]{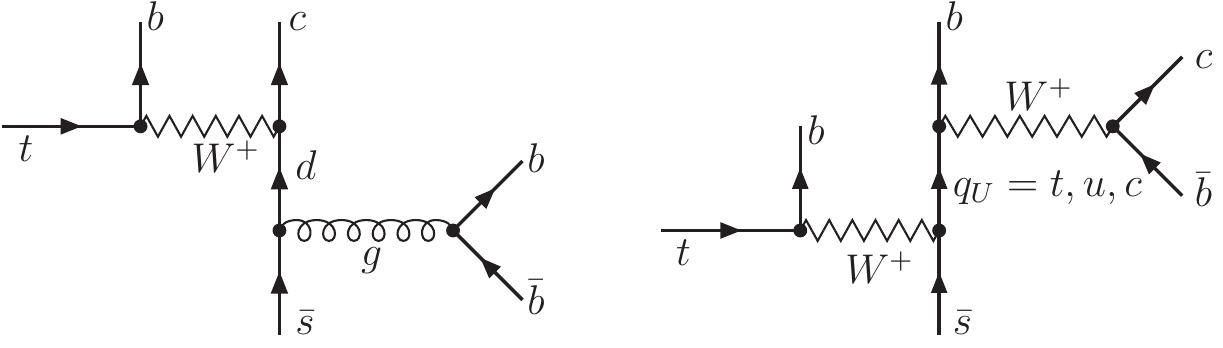}
\end{center}
\vspace{-5mm}
\caption{The representative diagrams  describing the contribution to
  $t \to b \bar{b} b \, c \bar{s}$
  decay, where the  $c \, \bar{s}$ pair is not from the $W$-boson.
   \label{fig:diagrams_csbar}
  }
\end{figure}

However, this is not true  for the $W$-boson decay
into quarks. For example,  there are additional  diagrams,
where the  $c \, \bar{s}$ pair  is produced not from $W$-boson (see
Fig.~\ref{fig:diagrams_csbar}).

\begin{figure}[ht!]
 \begin{center}
\includegraphics[width=0.6\textwidth,clip]{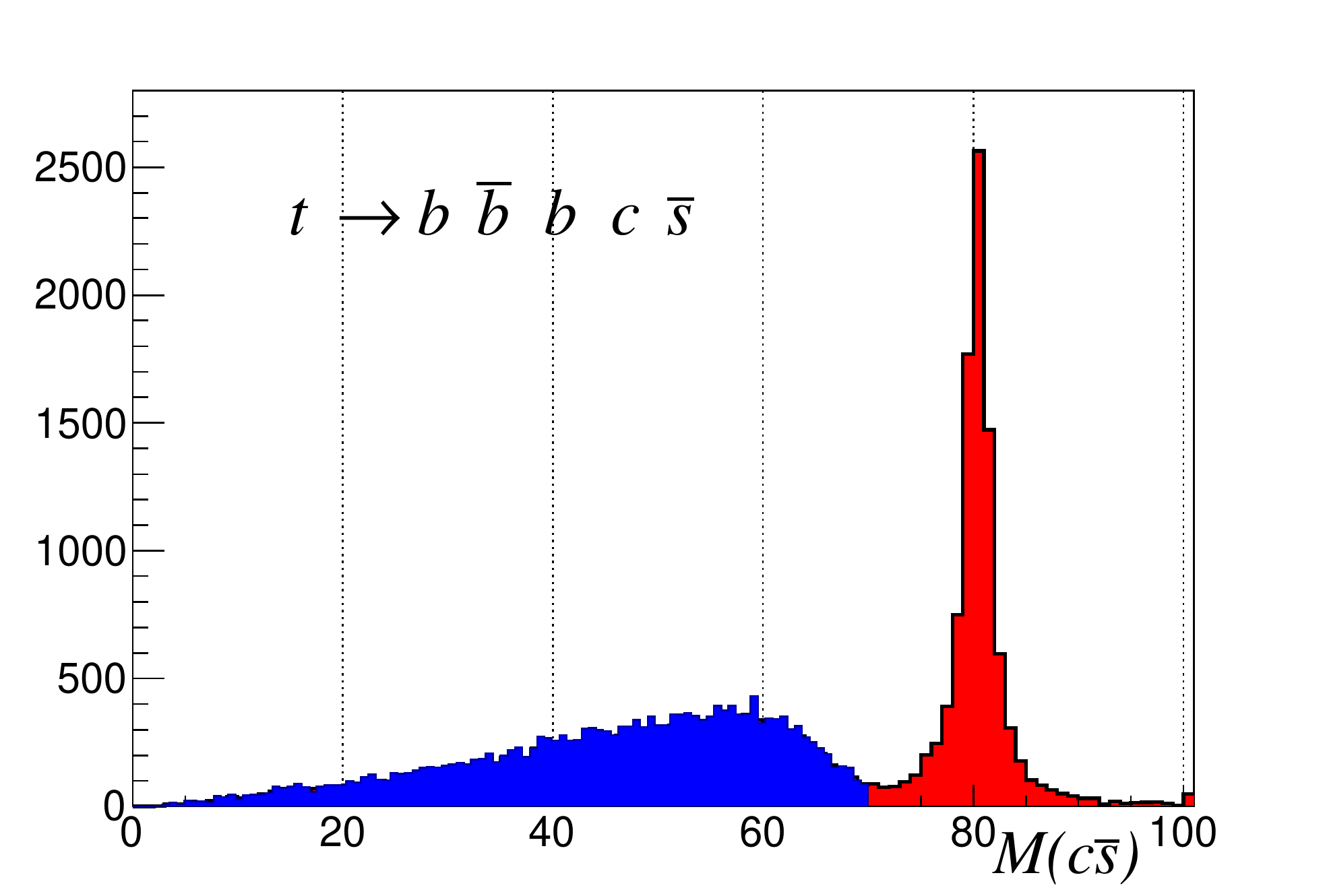} 
\end{center}
\vspace{-5mm}
\caption{The distribution over the invariant mass of the $c\bar{s}$ pair
  in the $t \to b \bar{b} b \, c \bar{s}$
  decay. The right sharp peak corresponds to virtual $W$-boson
  decay into $c\bar{s}$ pair, while the left wide region
  corresponds to the case when the  $s \, \bar{s}$ pair is not
  from the $W$-boson.
   \label{fig:minv_csbar}
  }
\end{figure}

\noindent Here we present few details concerning  top-quark decay channel:
$ t \to b \bar{b} b \, c \bar{s}$. 
The Fig.~\ref{fig:minv_csbar} presents the
distribution over the invariant mass of the $c\bar{s}$ pair
in the $t \to b \bar{b} b \, c \bar{s}$   decay.

As it follows from calculations, the contribution to the decay width
from this ``non-resonant'' (i.e. $c \bar{s}$ is  not from the $W*$-boson)
region approximately
coincides with the contribution from $W \to q \bar{q}'$ region.
As a result, the decay with of the 5-body top-quark decay channel
is approximately twice the $t \to b \bar{b} b \, W$ decay width: 
\begin{eqnarray}
  &&
  \left\{
  \begin{array}{lcl}
    \Gamma(t \to b \bar{b} b \, \ell^+ \nu) &=& (1.0 \pm 0.2)
    \times 10^{-4} \;\; \hbox{GeV}
    \\
    \Gamma(t \to b \bar{b} b \, c \bar{s})
    &=& (6.8 \pm 0.7) \times 10^{-4} \;\; \hbox{GeV}
  \end{array}
  \right.
  \nonumber
  \\
  \Rightarrow &&
  \Gamma(t \to b \bar{b} b \, f \bar{f}') =
  3\,\Gamma( \ldots \, \ell^+ \nu) +
  2 \, \Gamma(\ldots \,  c \bar{s}) 
  = (16.6 \pm 1.0)     \times 10^{-4} \;\; \hbox{GeV}
 \label{eq-11}
\end{eqnarray}
Therefore, the decay width value~(\ref{eq-11}) is about twice as large 
than $\Gamma(t \to b \bar{b} b \, W^+) = 9.3 \times 10^{-4}$~GeV
from~(\ref{eq-8}).

This circumstance should be taken into account when searching for
such a rare
decay of the top quark. Therefore, we think that the most efficient way is
the search for this decay in lepton mode
$t \, \to \, b \, \bar{b} \, b \, \ell^+ \nu$. 
% \label{eq-12}
%\end{eqnarray}
Therefore, in what follows the decay of the  $W$-boson into leptons
will be assumed in this article.

%%%%%%%%%%%%%%%%%%%%%%%%%%%%%%%%%%%%
\section{Top-quark decay $t \, \to \, \ell^+_1 \ell^-_1 \ell_2^+ \nu_2$ }
\label{fourleptons}

It was established many years ago, that
at tree level, the decay $t \, b \, Z \, W$  has some 
peculiar features, since the process occurs near the kinematic 
threshold
($m_t \sim M_Z + M_W +m_b$)~\cite{Decker:1992wz, Mahlon:1994us,
  Jenkins:1996zd, Altarelli:2000nt,
  Papaefstathiou:2017xuv, Onyisi:2017mza, Quintero:2014lqa}.
A brief discussion  of this decay channel can be found also
in~\cite{Beneke:2000hk}.
Note, that charged Higgs (at large $\tan \beta$ values)
should also  make a significant contribution to
the top-quark decay into four leptons:
\begin{eqnarray}
  t \, \to \, b \, Z^{*}(\ell^+_1 \ell^-_1)  \; W^{+*} ( \ell_2^+ \nu_2) \,
  \;\; + \;\; b \,  Z^{*}(\ell^+_1 \ell^-_1)
 \, \widetilde{H^{+*}} ( \ell_2^+ \nu_2)
 \label{eq-4lep}
\end{eqnarray}
where $ \widetilde{H^{+*}}$ corresponds to all charged Higgs contributions,
like
$t \to b \, H^{+ \, *} (\to \ell^+_1 \, \ell^-_1 \, \ell_2^+ \, \nu_2)$,
$t \to b Z^{*}  \; H^{+*} (\to  \ell_2^+ \nu_2)$, $\ldots$

Here, for the sake of completeness, we recalculate the width of this
decay. The results of calculations for the SM case and taking
into account the $\hbpm$-boson contribution, are given in the
Table~\ref{tab-4lep}. 
\begin{table}[h!] 
  \caption{
    The partial widths for
 $t \to b \, \ell^+_1 \, \ell^-_1 \, \ell_2^+ \, \nu_2$ decay.
     All width values (in GeV) are multiplied by $10^8$.
     }
  \label{tab-4lep}

  \begin{center}
 \renewcommand{\arraystretch}{1.2}
 \begin{tabular}{l||c|c|c|c}
   $M(H^{\pm}), \; \tan \beta$
  & $\mu^+ \mu^- e^+ \nu_e$ 
  & $\mu^+ \mu^- \tau^+ \nu_{\tau}$ 
  & $\tau^+ \tau^- \mu^+ \nu_{\mu}$ 
   & $\tau^+ \tau^- \tau^+ \nu_{\tau}$ 
  \\ \hline
  SM  & $6.1 \pm 0.3 $ &$6.1 \pm 0.3 $ & $6.1 \pm 0.3 $ & $5.7 \pm 0.24$
  \\ \hline \hline
  $160, \;\; 30$
  & $6.13 \pm 0.4$ & $11.8 \pm 0.9$ & $28. \pm 4.$ & $34.7 \pm 3.$
  \\ \hline 
  $170, \;\; 75$
  & $6.23 \pm 0.5$ & $7.6 \pm 0.5$ & $14.9 \pm 0.7$ & $17.4 \pm 0.6$
  \end{tabular}
\end{center}
\end{table}
\renewcommand{\arraystretch}{1.}

\noindent 
As can be seen from this table, for large values of $\tan\beta$
the  $\hbpm$-boson decay into $\tau$-leptons gives a noticeable
contribution to this partial width.
However, the branching fraction is too small to for 
experimental searches and will not be considered further.

%%%%%%%%%%%%%%%%%%%%%%%%%%%%%%%%%%%%
\section{Backgrounds 1. Four-body top-decays }
\label{background1}

In this section the main sources of the background processes
to the investigated decay $t \to b W^+ b \bar{b}$ are
discussed.  
We consider two types of background processes: top-quark decay
into other final states  and $t \bar{t}$ production
in hadronic collisions accompanied by others
partons (for example $p p \, \to \, t \bar{t} \, b \bar{b}$). 

We start with the  other four-body top-quark decay channels:
\begin{eqnarray}
  t \to b W^+ \, g g \;\; \hbox{and} \;\;
  t \to b W^+ \, q \bar{q}'  
  \label{eq-13}
\end{eqnarray}
There are about 30 decay channels with $(W \, q_1 \bar{q}_2 q_3)$
final state  and
three decay channels with gluons: $t \to q g g  W^+, \; q = d, s, b$.
The most part of these decays have very small decay widths. Therefore,
we present the results for such decays that have widths at least
10\%  of the  $t \to b \, W \, b \, \bar{b}$ decay width. The values
of the calculated widths  are given in the Table~\ref{tab-4}.

\begin{table}[h!] 
  \caption{
    The partial widths for $ t \to b \, W \, \bar{q}_1 q_2 (gg)$
    decay channels. The widths are in GeV.
    The values in third column  are evaluated
    with requirement that invariant mass of any two particle in the final
    state should be greater then 20 GeV.
    The fourth column (``ratio'')
    presents the ratios of the corresponding decay widths
    to the $\Gamma(b W b \bar{b})$ width, taking into account the
   $b$-tagging efficiency from~(\ref{eq-btag}). 
     }
  \label{tab-4}

  \begin{center}
 \renewcommand{\arraystretch}{1.2}
 \begin{tabular}{c||c|c|c}
$t \to b \, W^+ \, \ldots$ & no cuts & $M(i \, j) > 20$~GeV  & ratio
  \\ \hline
  $b \, \bar{b} $   & $(9.30 \pm 0.03) \times 10^{-4}$
                       & $(1.34 \pm 0.03) \times 10^{-4}$ & 1
   \\ \hline 
   $ g \, g $   & &  $(1.8 \pm 0.04) \times 10^{-3}$ & 0.003
\\ 
\hline
$c \, \bar{c} $   & $(4.2 \pm 0.3) \times 10^{-3}$
                    &  $(1.3 \pm 0.3) \times 10^{-4}$  & 0.02
\\   
  $s \, \bar{s} $   & $(1.1 \pm 0.15) \times 10^{-2}$
                    & $(1.3 \pm 0.3) \times 10^{-4}$ & 0.0002
  \\   
  $d \bar{d} \; (u \, \bar{u}) $
                    & $(1.6 \pm 0.4) \times 10^{-2}$
                    &  $(1.3 \pm 0.3) \times 10^{-4}$ & 0.0002
  \end{tabular}
\end{center}
\end{table}
\renewcommand{\arraystretch}{1.}
As it seen from third column of this table the decay widths of
these channels
are comparable (or even greater) then the width
of the main process 
 $\Gamma(t \to b \bar{b} b \, W^+) = 9.3 \times 10^{-4}$~GeV
from~(\ref{eq-8}). However, the $b$-tagging application
provides a substantial suppression of the background decay
channels (see fourth column in the Table~\ref{tab-4}).

%%%%%%%%%%%%%%%%%%%%%%%%%%%%%%%%%%%%
\section{Backgrounds 2. The production processes }
\label{background2}
In this section we consider the backgrounds
from $t \bar{t}$ hadronic production
 accompanied by two  partons:
 \begin{eqnarray}
 %  \left.
 %  \begin{array}{l}
     p\, p \, \to \, t \, \bar{t} \, j_1 j_2,
     \quad j_1 j_2  = b \, \bar{b}, \;\; g\,g, \;\; g\,q,
     \;\; q \, \bar{q}' 
\label{eq-ttjj}  
\end{eqnarray}
 As it shown latter the main background comes from $t\bar{t} b \bar{b}$
 production in $pp$-collisions. For $pp$-collisions at $\sqrt{s} = 13$~TeV
 the cross-section values  were calculated~\cite{Bredenstein:2008zb,
   Cascioli:2013era, Garzelli:2014aba,
 Bevilacqua:2011aa, Bevilacqua:2014qfa, Bevilacqua:2017cru, Jezo:2018yaf}
 and were measured by the ATLAS and CMS
 Collaborations~\cite{ATLAS:2018fwl,CMS:2020grm,CMS:2019eih}.

 Following these results (for $t \bar{t} b \bar{b}$)
for further calculations  we use the values for  cross section production
of $t \bar{t}$ reaction
($\sigma(t \bar{t}) = 803 \pm 2 (\hbox{stat}) \pm 25
(\hbox{syst}) \pm 20(\hbox{lumi})$~pb (see~\cite{Grancagnolo:2019uvn}
and reference therein) and
for $t \bar{t} b \bar{b}$ process 
\begin{eqnarray}
  \left.
  \begin{array}{lcl}
  \sigma(t \bar{t}) &=& 800  \;\;\; \hbox{pb}
    \\
  \sigma (t \bar{t} b \bar{b}) & = & 4  \;\;\; \hbox{pb}
   \end{array}
  \right\}
     \label{eq-sigmas}
\end{eqnarray}

The evaluated cross-section production values for reactions~(\ref{eq-ttjj})
at $\sqrt{s} = 13$~TeV are given in Table~\ref{tab-5} below.
These values are evaluated with the following kinematic cuts:
\begin{eqnarray}
  p_{\top}(j) \ge 20 \; \;\hbox{GeV},
  \quad M(j_1 j_2) \ge 20\;\; \hbox{GeV}
  \label{eq-cuts}
  \end{eqnarray}
As it seen the requirement to have two $b$-tagged jets
($jj \to$ ($b$-tagging) $\to j_b j_b$)
provides the substantial suppression of the all background processes
except the $p p \to t \bar{t} b \bar{b}$ production.

\begin{table}[h!] 
  \caption{
 The cross section values for the processes $p p \to t \bar{t} j_1 j_2$
 (where $j$ stands for quarks or gluon) with
 kinematic cuts $p_{\top}(j) \ge 20$~GeV and $M(j_1 j_2) \ge 20$~GeV.
 The third column presents cross section values times $b$-tagging
 efficiencies: $\epsilon_b = 0.7, \epsilon_c = 0.1,
 \epsilon_{q, \; g} = 0.01$ from~(\ref{eq-btag}).
     }
  \label{tab-5}
\begin{center}
 \renewcommand{\arraystretch}{1.2}

 \begin{tabular}{l|c|c||c}
   {\it reaction}
   & $\sigma$~(pb) & $\sigma \times \epsilon_1 \epsilon_2$~(pb)
   & $\sigma / \sigma(t \bar{t} \, b \bar{b})$\\ \hline
 $p p \to t \, \bar{t} \, b \, \bar{b}$ & $0.76 \pm 0.08$
 &  $0.372 \pm 0.040$  & $1$ 
 \\ \hline \hline
 $p p \to t \, \bar{t} \, g \, g$ & $57.3 \pm 12.8$
 & $(5.73 \pm 1.28) \times 10^{-3}$ & $1.5 \times 10^{-2}$
 \\ \hline
$ p p \to t \bar{t} u \bar{u} $ & $0.61 \pm 0.08$ &
  $(6.1 \pm 0.8) \times 10^{-5}$& $1.6 \times 10^{-3}$ 
 \\  \hline
$p p \to t \bar{t} u u$ & $0.64 \pm 0.28$ & $(6.4 \pm 2.8) \times 10^{-5}$
 & $1.7 \times 10^{-4}$
 \\
 \hline
$ p p \to t \bar{t} d \bar{d} $ & $0.53 \pm 0.08$ & 
 $(5.3 \pm 0.8) \times 10^{-5}$ & $1.4 \times 10^{-3}$ 
 \\  \hline
 $p p \to t \bar{t} d d$ & $0.26 \pm 0.2$ & $(2.6 \pm 2.) \times 10^{-5}$
 & $7.0 \times 10^{-5}$
 \\ \hline
 $p p \to t \, \bar{t} \, c \, \bar{c} $ & $0.56 \pm 0.11$
 & $(5.6 \pm 1.0) \times 10^{-3}$ & $1.5 \times 10^{-2}$
 \\ \hline
  $ p p \to t \bar{t}\, b \,g $ & $0.37 \pm 0.14$ &
 $(3.7 \pm 0.14) \times 10^{-3}$ & $1.0 \times 10^{-2}$ 
 \\  \hline
 $ p p \to t \bar{t}\, \bar{b} \, g $ & $0.44 \pm 0.12$ &
 $(4.4 \pm 1.2) \times 10^{-3}$ & $1.2 \times 10^{-2}$ 
  
       \end{tabular}
\end{center}
\end{table}
\renewcommand{\arraystretch}{1.}
Thus, the main background process comes from the $t \bar{t} b \bar{b}$
production in $pp$-collisions. This background can be suppressed by
applying  kinematic cuts to the  final partons. Indeed,
the processes of $t \bar{t}$ production with the subsequent four-body
top-quark decay and $t \bar{t} b \bar{b}$ production have the same final
state:
\begin{eqnarray}
  && p\, p \, \to \, t \, (\to \, b \bar{b} b W^+) 
  \;\; \bar{t} (\to \, \bar{b} W^-)
  \;\; \;\; \;\; \to b \bar{b} b W^+ \bar{b} W^-
    \label{eq-tt4}
    \\
    &&   p\, p \, \to \, t \, (\to \, b  W+)
    \; \;\; \; \;  \bar{t} \, (\to \, \bar{b} W^-) \;
     b \bar{b}  \; \to b \bar{b} b W^+ \bar{b} W^-
\label{eq-ttbb}  
\end{eqnarray}

In what follows the $b$-quark and $\bar{b}$-quark will be
considered as a $b$-jet.
Therefore, the reactions have to following (identical) final states:
$4 j_b W_1 W_2$.
The pair $j_b$ and $W$, with a mass closest to the mass of the top quark
is treated  as a top-quark decaying into $b$ and $W$.
The system system consisting of the remaining particles
(three $b$-jets and $W$), is considered as
``top''-quark $T_{3b}$.
Inside this $T_{3b}$-system we select a pair of $j_b$ and $W$
with  maximum mass. This pair is treated as a ``virtual top-quark''
($t^*$).

\begin{figure}[ht!]
\begin{center}
\includegraphics[width=0.45\textwidth,clip]{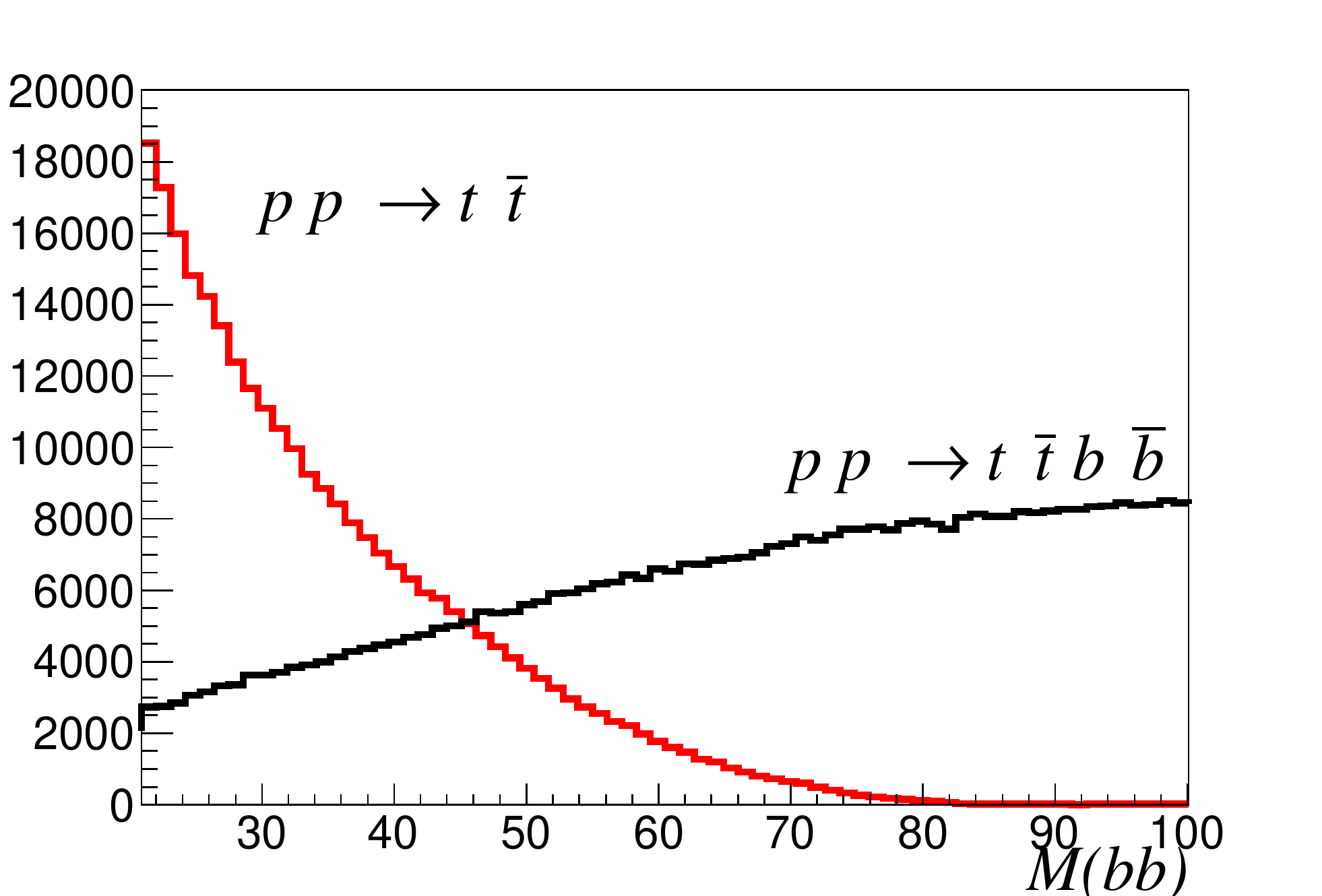} 
\includegraphics[width=0.45\textwidth,clip]{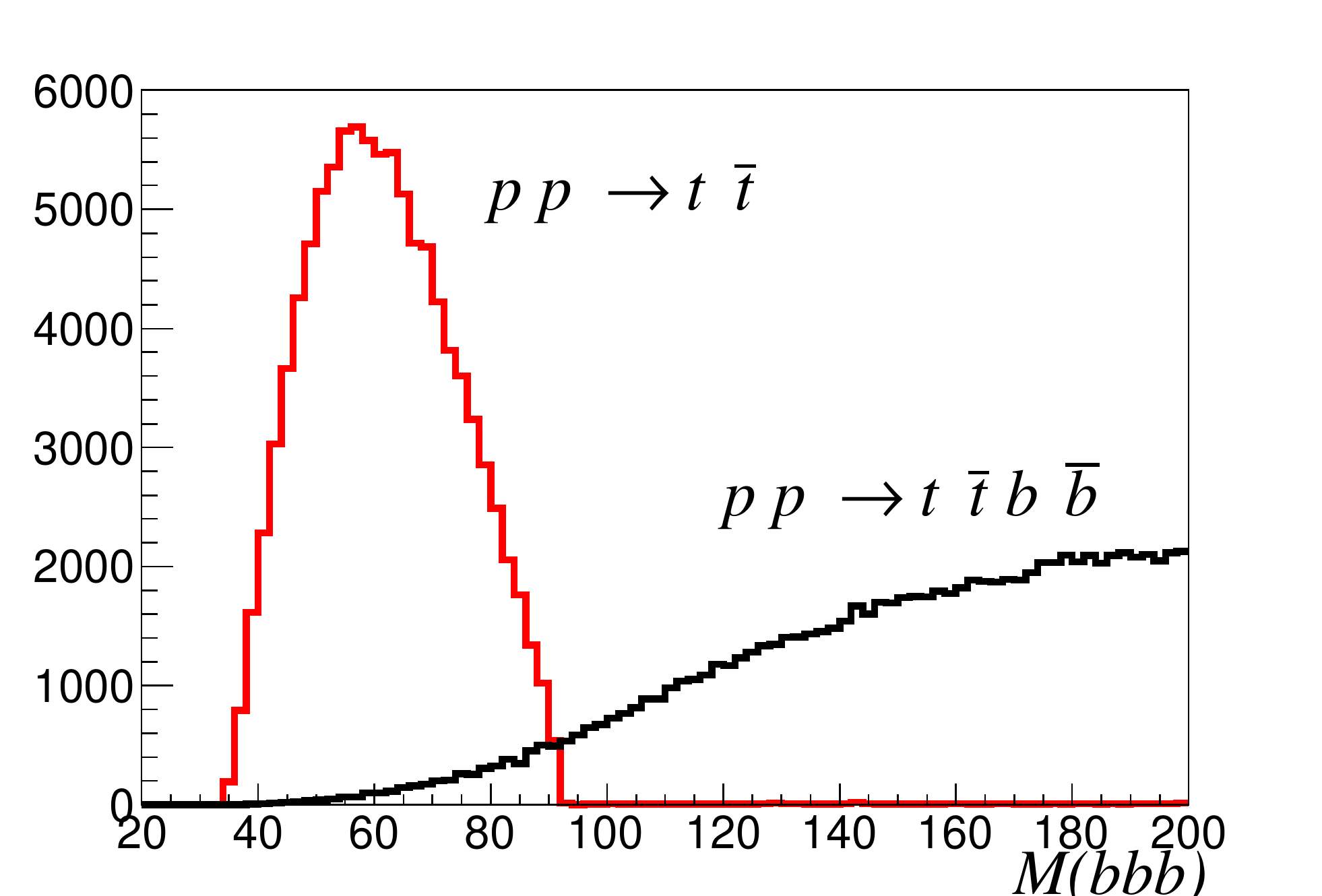} 
\\
\includegraphics[width=0.45\textwidth,clip]{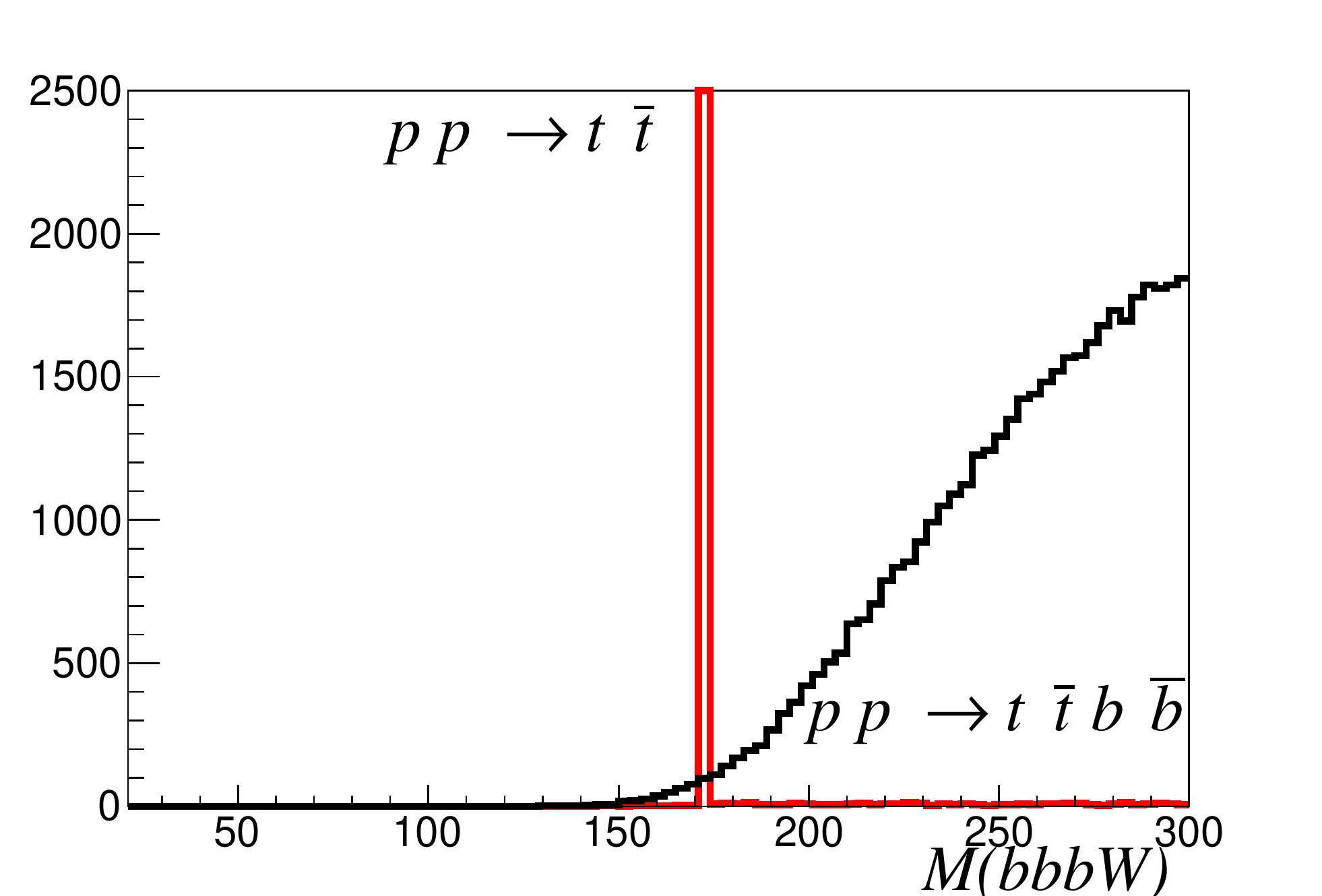} 
\includegraphics[width=0.45\textwidth,clip]{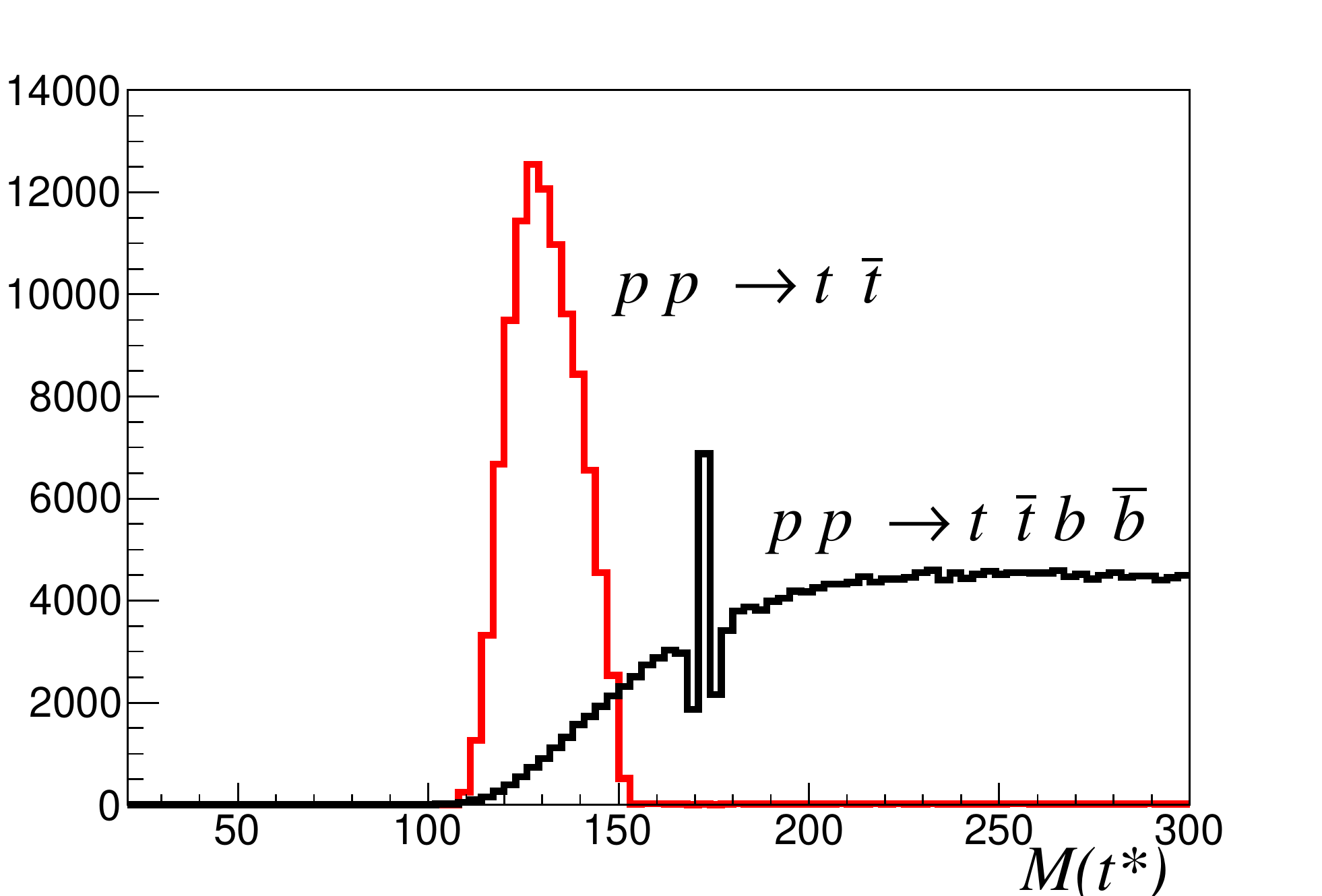} 
\end{center}
\vspace{-5mm}
\caption{$d \, N / M_{inv}$ distributions. The mass of two (three)
  $b$-quarks
is shown in top left (top right) plot. 
The distributions over $3 j_b \, W$ (virtual top $t^*$)
is shown in the bottom left (bottom right) plot.
    \label{fig:minv_pairs}
  }
\end{figure}

Note, that the distributions over the invariant masses of two and three
$b$-quarks and the virtual top mass are very
different for the reactions~(\ref{eq-tt4})
and~(\ref{eq-ttbb}). The final state $3 j_b W$ in the
reaction~(\ref{eq-tt4})  has a sharp
peak  (due to top-quark decay).
These distributions are presented in the Figs.~\ref{fig:minv_pairs}.

Therefore, to suppress the background from
$t \bar{t} b \bar{b}$ production~(\ref{eq-ttbb}) we apply the
kinematic cuts as follows:
\begin{eqnarray}
  \left.
  \begin{array}{lcclcc}
  M(j_1 \, j_2) & \ge& 20 \;\; \hbox{GeV}, & 
  p_{\top}(j) &\ge& 20 \;\; \hbox{GeV}
  \\
 M(j_b \, j_b) & \le & 60 \;\; \hbox{GeV}, & 
  M(3 j_b) & \le & 90 \;\; \hbox{GeV}
  \\
  M(3 j_b \, W) & \le&  200 \;\; \hbox{GeV}, & 
  M(t^*)  &=& 100 \div 150 \;\; \hbox{GeV}
  \end{array}
  \right\}
  \label{eq-cuts-all}
\end{eqnarray}
After applying of all  cuts from~(\ref{eq-cuts-all}) the
total efficiencies ($\varepsilon_{cuts}$) of event detection
for the $T_{3b}(b \bar{b} b W)$ system are as follows:
\begin{eqnarray}
    \left.
  \begin{array}{lcl}
    p p \to t \bar{t}  & :          & \varepsilon_{cuts} = 0.086
    \\
    p p \to t \bar{t} b \bar{b} & : &\varepsilon_{cuts} = 0.00034
  \end{array}
  \right\}
  \label{eq-rescuts}
\end{eqnarray}

As a result   the application of all cuts~(\ref{eq-cuts-all})
provides a rather well suppression of the background process:
\begin{eqnarray}
  R = \frac{\sigma(t \bar{t} b \bar{b})\times \epsilon_{cuts}}
  {\sigma(t \bar{t} )\times \epsilon_{cuts} \times \Br(t \to 3b W)}
    = 0.031
   \label{eq-rsigma} 
  \end{eqnarray}
where $\sigma(t \bar{t} b \bar{b})$ and $\sigma(t \bar{t})$ are
taken from~(\ref{eq-sigmas}).
Therefore, in what follows we consider the signal
process~(\ref{eq-tt4}) only.

\noindent 
Then, for estimation the expected number of events
we use the following options:
\\[1mm]
$\bullet$
the LHC Run-3 integrated luminosity equals $L_{tot} = 300$~fb${}^{-1}$;
\\
$\bullet$ $\Br(t \to b \bar{b} b W) = 6.29 \times 10^{-4}$;  \\
$\bullet$ the kinematic cuts from~(\ref{eq-cuts-all});
\\
$\bullet$ $W^+ W^-$ decay into $\ell \, \nu$ and $q \bar{q}'$ : 
$W^+ W^- \to e(\mu) \, \nu \;\; q \bar{q'}$. 

As a result, at LHC Run-3 option the expected  number
of events for $t \bar{t}$-pair
production with subsequent decays (with the following $W^+W^-$ decays)
are as follows:
\begin{eqnarray}
  && t \, \bar{t} \to b \bar{b} b W^+ \, \bar{b} W^-
  \quad \to \;\;  2b \, 2 \bar{b} \, \ell \nu \, q \, \bar{q}'
  \nonumber
  \\
 &&  N(2b \, 2 \bar{b} \, \ell \nu \, q \, \bar{q}') \approx 2000  
  \label{eq-16}
\end{eqnarray}
This number of events is large enough that the search for such 
rare decay looks like a promising goal for the experimental searches.

\section {Kinematics}
\label{kinematics}

In this section we study the obvious question: {\it
  how one get a  manifestation of the charged Higgs boson
  contribution to this rare $t \to b W^+ b \bar{b}$ decay of the
  top-quark ?}.

In this section , the symbol "SM" corresponds to the case when
top-quark decays within the SM,  while "H" denotes the inclusion
(exchange) of
 charged Higgs boson:
\begin{eqnarray}
  t_{SM} \to b \bar{b} b \, W^+ &:& \quad \hbox{(SM \;only)}
  \label{eq-dec-sm}
  \\
   t_{H} \to b \bar{b} b \, W^+ &:& \quad \hbox{(SM}
  \; + \; \hbpm) 
  \label{eq-dec-h}
\end{eqnarray}

As it seen from the Table~\ref{tab-3} for a small $\tan \beta$ values
$(\tan \beta \le 0.2)$ the  decay width due to $H^{\pm}$
contribution exceeds the SM value by at least one
(or even two) order of magnitude.

\begin{figure}[ht!]
\begin{center}
\includegraphics[width=0.45\textwidth,clip]{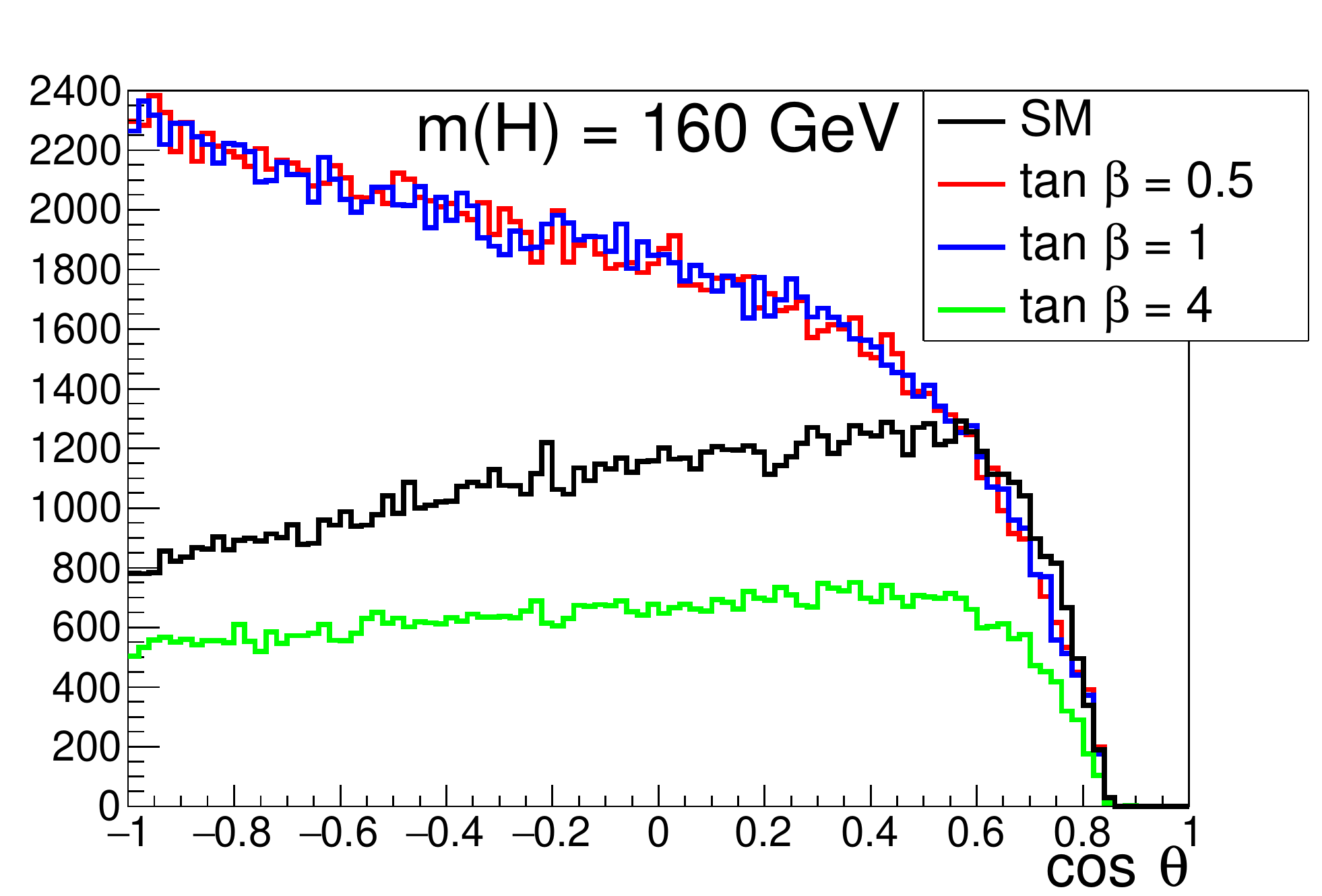}
\includegraphics[width=0.45\textwidth,clip]{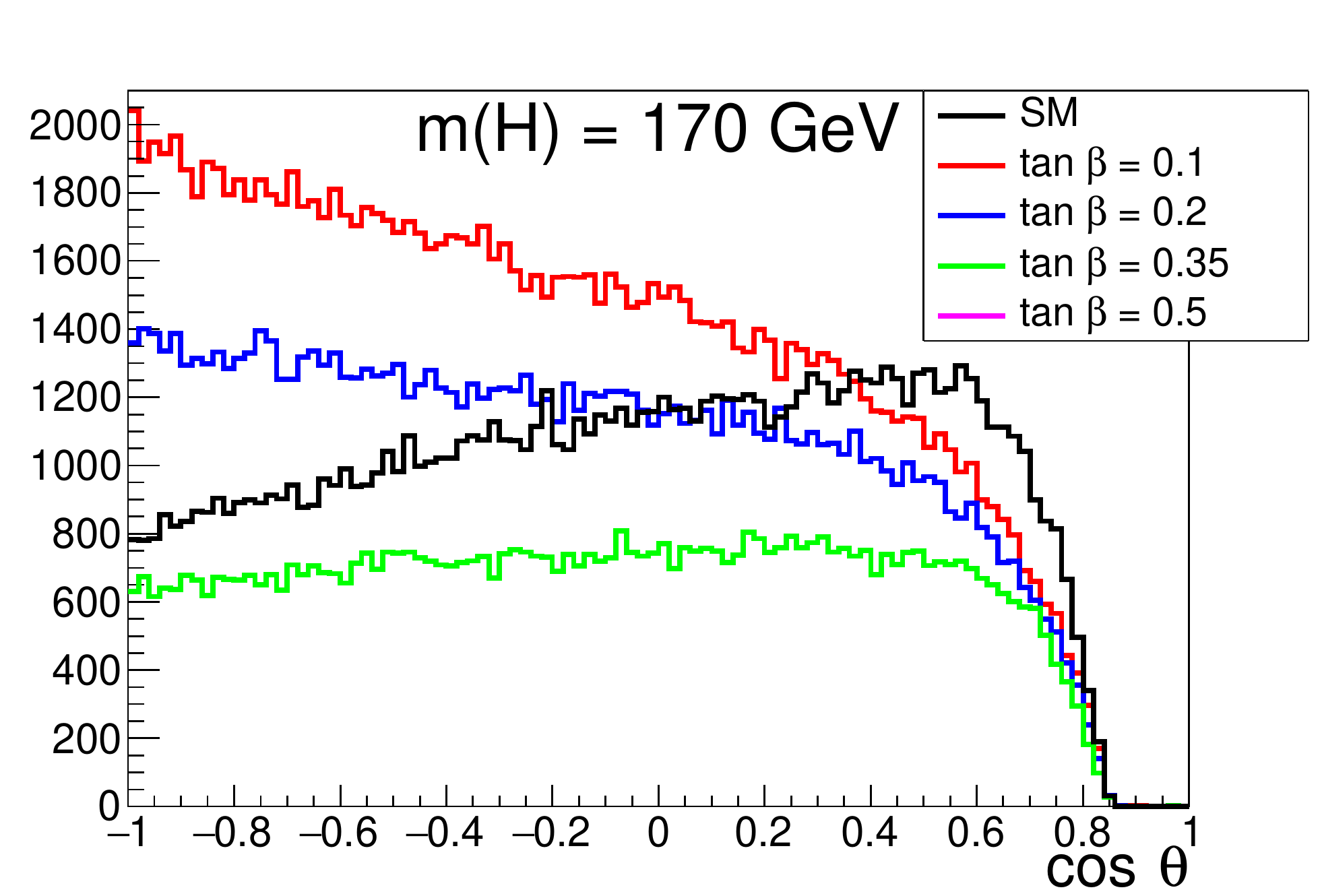}
\end{center}
  \vspace{-5mm}
  \caption{The distribution over $\cos \theta_{bb}$ for
    two charged Higgs masses and for several $\tan \beta$ values.
 The black curve corresponds to SM $\decbw$ decay. 
   \label{fig:cos_rest1}
  }
\end{figure}

\begin{figure}[ht!]
\begin{center}
\includegraphics[width=0.45\textwidth,clip]{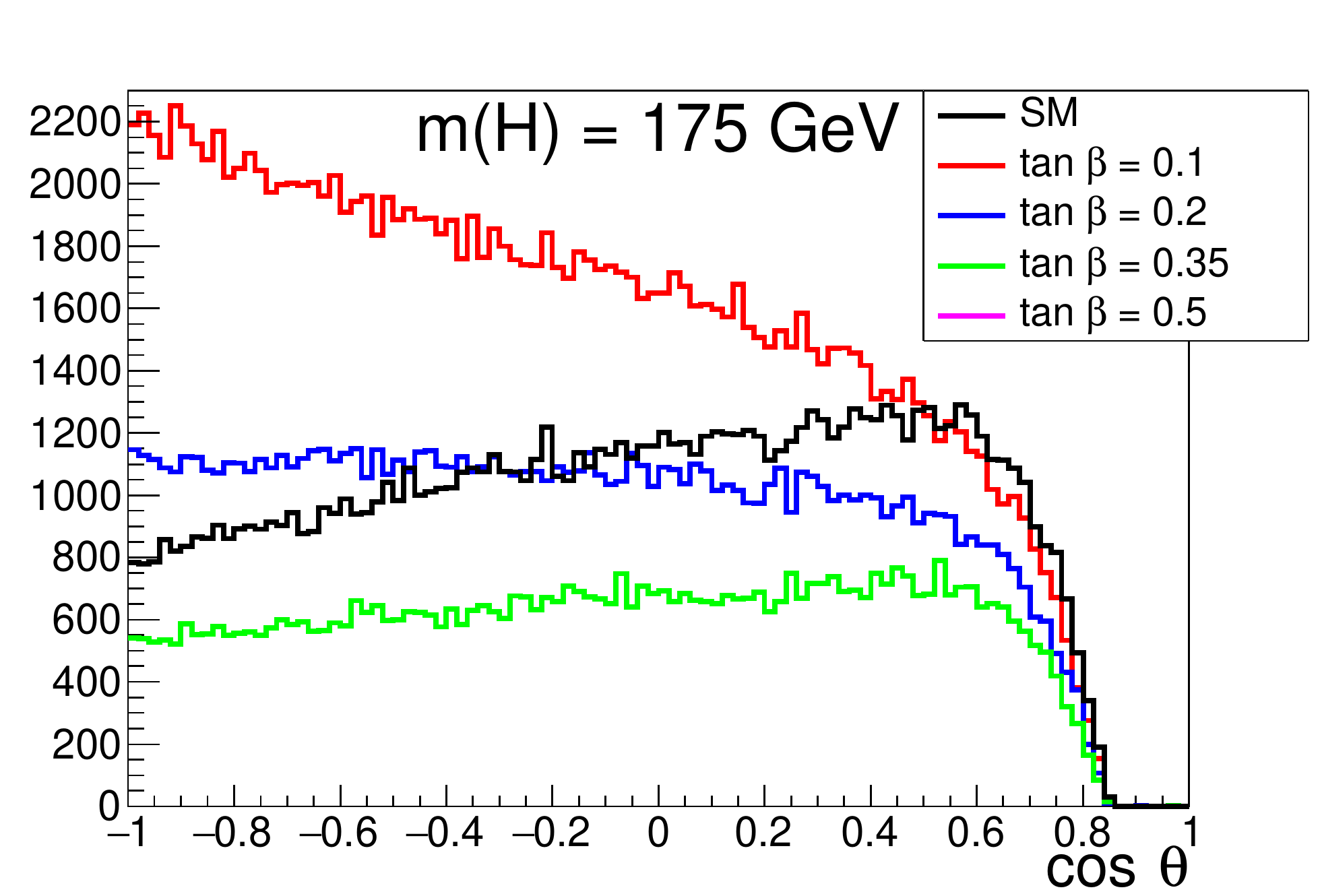}
\includegraphics[width=0.45\textwidth,clip]{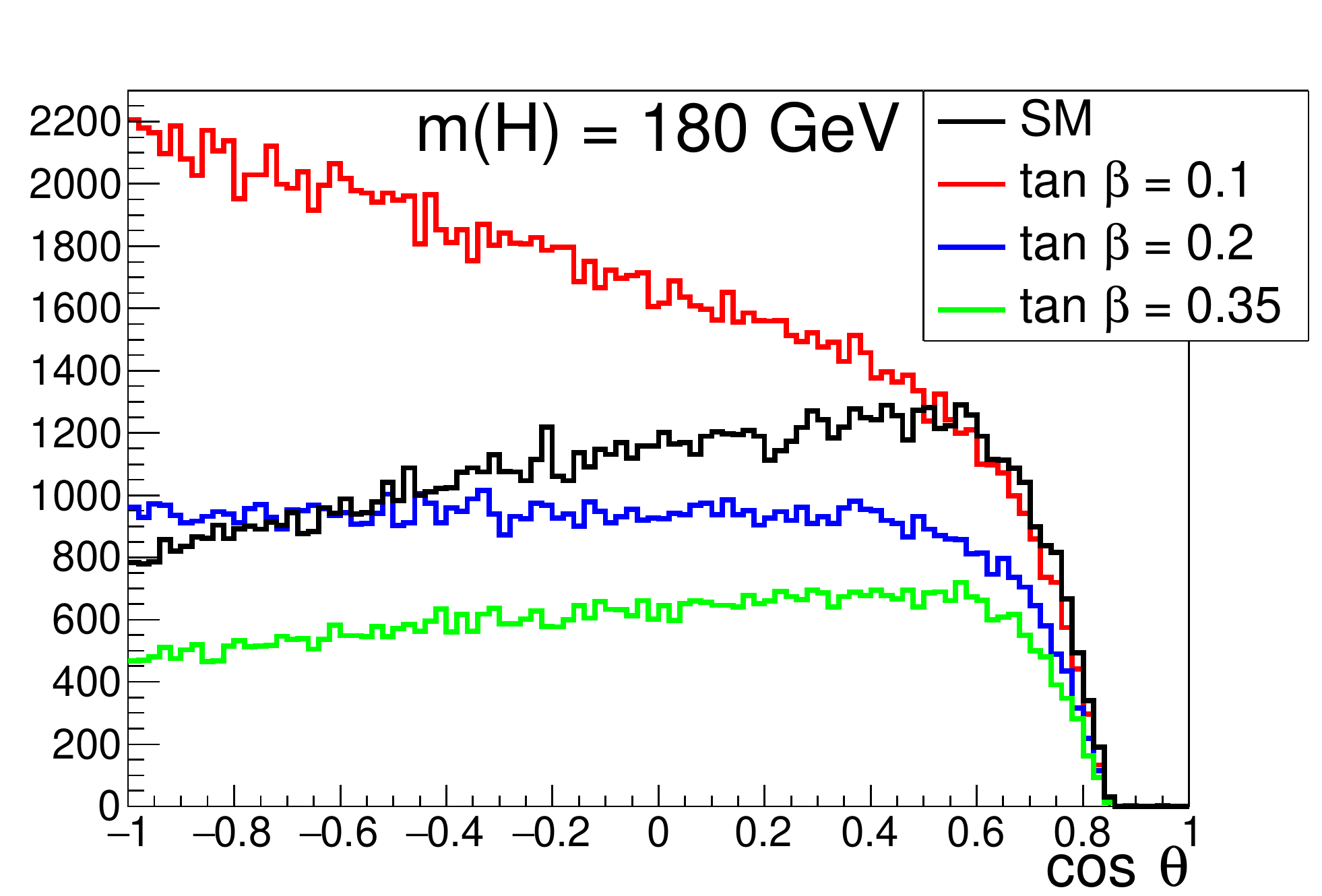}
\end{center}
  \vspace{-5mm}
  \caption{The  same distributions as in Fig.~\ref{fig:cos_rest1},
    but for $M(\hbpm) = 175$ and $180$~GeV.  
   \label{fig:cos_rest2}
  }
\end{figure}

Therefore, one could expect a significant increase in the number of events
in this case.
Additional manifestation of the $H^{\pm}$ contribution can be seen in
different kinematic distributions.
We have found one variable, whose distribution demonstrates 
 clear difference in decays  (\ref{eq-dec-sm})
and (\ref{eq-dec-h}).
This is  the cosine of the angle
between the 3-momenta of the
final $b$-quarks in the $t$-quark rest frame, $\cos \theta_{bb}$.
These distributions  are shown in the
Figs.~\ref{fig:cos_rest1} and~\ref{fig:cos_rest2}.

As can be seen from these plots, for small $\tan \beta$ values
($\tan \beta \le 0.2$)
these distributions for  $t_{H}$ decay~(\ref{eq-dec-h})
have a shape different from the SM decay.

\begin{figure}[ht!]
\begin{center}
\includegraphics[width=0.49\textwidth,clip]{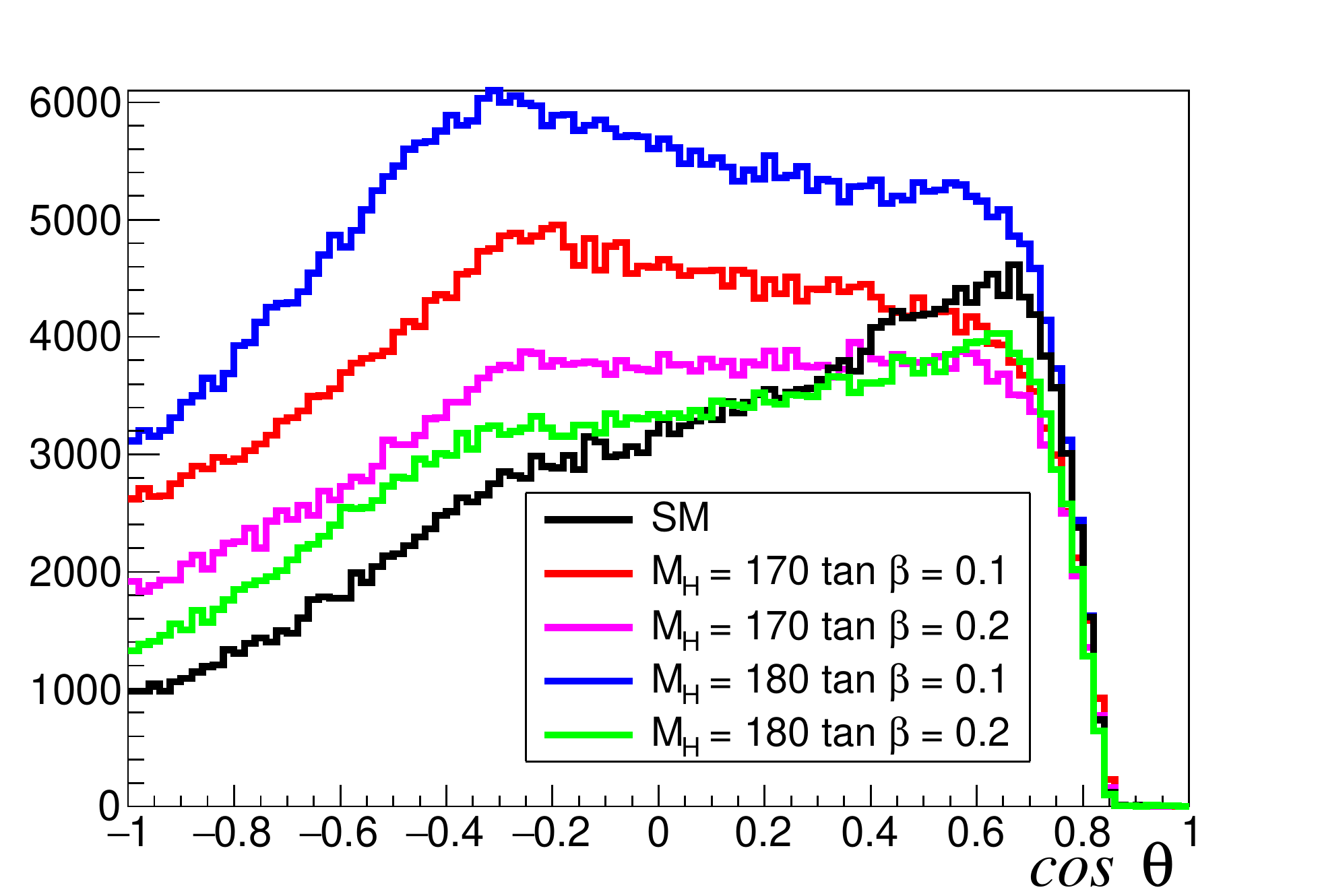} 
 \end{center}
  \vspace{-5mm}
  \caption{The distribution over $\cos \theta_{bb}$ for
    for-body top-quark decay. Here the $t$-quark is produced
    in $p p$-collisions at $\sqrt{s} = 13$~TeV.
    The black curve corresponds to  SM decay $\decbw$. 
   \label{fig:cos_pp}
  }
\end{figure}

The next plots (see Fig.~\ref{fig:cos_pp}) present the same distributions,
but evaluated for top-quark production in
$p p$-collisions at $\sqrt{s} = 13$~TeV after application all
cuts from~(\ref{eq-cuts-all}).
It is seen an obvious difference in  distributions over $\cos \theta_{bb}$.

Certainly, this conclusion is nothing more than a demonstration
of different kinematic in (\ref{eq-dec-sm}) and (\ref{eq-dec-h})
decays.
More realistic simulation could destroy this pictures and conclusions.
Nevertheless,  we believe that advanced methods of the
 experimental data analysis will make it possible 
 to search for this rare $t$-quark four-body decay
 and (in principle) to detect 
 manifestation of a charged Higgs boson (if it exists) to this
 $t \to b W^+ b \bar{b}$ decay.

%%%%%%%%%%%%%%%%%%%%%%%%

\section{Conclusion}
\label{conclusion}
In this article the rare top decay $\decbw$ is considered.
The relatively large partial decay width,  
$\Gamma(\decbw) = (9.30 \pm 0.03) \times 10^{-4}$~GeV
(and $\Br(\decbw) = 6.29 \times 10^{-4}$),  makes possible the searches
for this rare $t$-quark decay at LHC.

The role of the charged Higgs boson
(with the mass $M(\hbpm) = (160 \div 180)$~GeV and $\tan \beta \lsim 1$)
in top-quark decays is considered.
Taking in account the current value of the $t$-quark total decay width
as well as the branching fractions of $t$-quark decays, 
the allowable  region of $\tan \beta$ versus $M(\hbpm)$
is evaluated.

In this article we propose  an additional  method
for $\hbpm$-boson searches. This scenario explores only
the  increase of the partial width of the $\decbw$ decay 
 and does not require the appearance of the peak in the invariant
mass distributions around the mass of the charged Higgs boson.

It is shown that the $\hbpm$-boson  contribution to $\Gamma(\decbw)$
with $\tan \beta \lsim 1$
could increase this width by up to two orders of magnitude (depending
on  $\tan \beta$ values).

The role of the possible background processes
(for $t$-quark decay as well as for production) is investigated.
The exploration of the $b$-tagging ($\varepsilon_b \approx 70\%$)
provides sufficiently good  suppression of all other decay channels.
The main background to $t \bar{t}$ production with
subsequent $\decbw$ decay comes from four heavy quarks production
($pp \to t \bar{t} b \bar{b})$. This background also could be
suppressed by exploring of $b$-tagging and proper kinematic cuts.

As a result one may expect to get about 2000 events
 at LHC energies with total luminosity of ${\Lgr}_{tot} = 300$~fb${}^{-1}$
 for the SM $t$-quark decay $\decbw(\to e/\mu \, \nu)$.
  Such a number of expected events makes it realistic
challenge for this study is in the LHC experiments.

\vspace{10mm}
\section*{Acknowledgments}
\noindent
In conclusion the author is  grateful to P.S.~Mandrik, V.F.~Obraztsov,
R.N.~Rogalyov, and
A.M.~Zaitsev  for multiple and fruitful discussions.

%% file: Top3bw.bbl
\begin{thebibliography}{10}
\expandafter\ifx\csname url\endcsname\relax
  \def\url#1{\texttt{#1}}\fi
\expandafter\ifx\csname urlprefix\endcsname\relax\def\urlprefix{URL }\fi
\expandafter\ifx\csname href\endcsname\relax
  \def\href#1#2{#2} \def\path#1{#1}\fi

\bibitem{Beneke:2000hk}
M.~Beneke, et~al., {Top quark physics}, in: {Workshop on Standard Model Physics
  (and more) at the LHC (First Plenary Meeting)}, 2000, pp. 419--529 (3 2000).
\newblock \href {http://arxiv.org/abs/hep-ph/0003033}
  {\path{arXiv:hep-ph/0003033}}.

\bibitem{Zyla:2020zbs}
P.~A. Zyla, et~al., {Review of Particle Physics}, PTEP 2020~(8) (2020) 083C01,
  \url{https://pdg.lbl.gov/} (2020).
\newblock \href {https://doi.org/10.1093/ptep/ptaa104}
  {\path{doi:10.1093/ptep/ptaa104}}.

\bibitem{Barger:1987nn}
V.~D. Barger, R.~J.~N. Phillips, {COLLIDER PHYSICS}, Vol.~71, Addison-Wesley,
  1987 (1987).

\bibitem{Branco:2011iw}
G.~C. Branco, P.~M. Ferreira, L.~Lavoura, M.~N. Rebelo, M.~Sher, J.~P. Silva,
  {Theory and phenomenology of two-Higgs-doublet models}, Phys. Rept. 516
  (2012) 1--102 (2012).
\newblock \href {http://arxiv.org/abs/1106.0034} {\path{arXiv:1106.0034}},
  \href {https://doi.org/10.1016/j.physrep.2012.02.002}
  {\path{doi:10.1016/j.physrep.2012.02.002}}.

\bibitem{Akeroyd:2016ymd}
A.~G. Akeroyd, et~al., {Prospects for charged Higgs searches at the LHC}, Eur.
  Phys. J. C 77~(5) (2017) 276 (2017).
\newblock \href {http://arxiv.org/abs/1607.01320} {\path{arXiv:1607.01320}},
  \href {https://doi.org/10.1140/epjc/s10052-017-4829-2}
  {\path{doi:10.1140/epjc/s10052-017-4829-2}}.

\bibitem{Arbey:2017gmh}
A.~Arbey, F.~Mahmoudi, O.~Stal, T.~Stefaniak, {Status of the Charged Higgs
  Boson in Two Higgs Doublet Models}, Eur. Phys. J. C 78~(3) (2018) 182 (2018).
\newblock \href {http://arxiv.org/abs/1706.07414} {\path{arXiv:1706.07414}},
  \href {https://doi.org/10.1140/epjc/s10052-018-5651-1}
  {\path{doi:10.1140/epjc/s10052-018-5651-1}}.

\bibitem{ALEPH:2013htx}
G.~Abbiendi, et~al., {Search for Charged Higgs bosons: Combined Results Using
  LEP Data}, Eur. Phys. J. C 73 (2013) 2463 (2013).
\newblock \href {http://arxiv.org/abs/1301.6065} {\path{arXiv:1301.6065}},
  \href {https://doi.org/10.1140/epjc/s10052-013-2463-1}
  {\path{doi:10.1140/epjc/s10052-013-2463-1}}.

\bibitem{CDF:2009esh}
T.~Aaltonen, et~al., {Search for Higgs bosons predicted in two-Higgs-doublet
  models via decays to tau lepton pairs in 1.96-TeV p anti-p collisions}, Phys.
  Rev. Lett. 103 (2009) 201801 (2009).
\newblock \href {http://arxiv.org/abs/0906.1014} {\path{arXiv:0906.1014}},
  \href {https://doi.org/10.1103/PhysRevLett.103.201801}
  {\path{doi:10.1103/PhysRevLett.103.201801}}.

\bibitem{D0:2011rhz}
V.~M. Abazov, et~al., {Search for Higgs bosons of the minimal supersymmetric
  standard model in $p\bar{p}$ collisions at $\sqrt{s}=1.96$ TeV}, Phys. Lett.
  B 710 (2012) 569--577 (2012).
\newblock \href {http://arxiv.org/abs/1112.5431} {\path{arXiv:1112.5431}},
  \href {https://doi.org/10.1016/j.physletb.2012.03.021}
  {\path{doi:10.1016/j.physletb.2012.03.021}}.

\bibitem{ATLAS:2018fwl}
M.~Aaboud, et~al., {Measurements of inclusive and differential fiducial
  cross-sections of $ t\overline{t} $ production with additional heavy-flavour
  jets in proton-proton collisions at $ \sqrt{s} $ = 13 TeV with the ATLAS
  detector}, JHEP 04 (2019) 046 (2019).
\newblock \href {http://arxiv.org/abs/1811.12113} {\path{arXiv:1811.12113}},
  \href {https://doi.org/10.1007/JHEP04(2019)046}
  {\path{doi:10.1007/JHEP04(2019)046}}.

\bibitem{ATLAS:2018gfm}
M.~Aaboud, et~al., {Search for charged Higgs bosons decaying via $H^{\pm} \to
  \tau^{\pm}\nu_{\tau}$ in the $\tau$+jets and $\tau$+lepton final states with
  36 fb$^{-1}$ of $pp$ collision data recorded at $\sqrt{s} = 13$ TeV with the
  ATLAS experiment}, JHEP 09 (2018) 139 (2018).
\newblock \href {http://arxiv.org/abs/1807.07915} {\path{arXiv:1807.07915}},
  \href {https://doi.org/10.1007/JHEP09(2018)139}
  {\path{doi:10.1007/JHEP09(2018)139}}.

\bibitem{ATLAS:2021upq}
G.~Aad, et~al., {Search for charged Higgs bosons decaying into a top quark and
  a bottom quark at $ \sqrt{\mathrm{s}} $ = 13 TeV with the ATLAS detector},
  JHEP 06 (2021) 145 (2021).
\newblock \href {http://arxiv.org/abs/2102.10076} {\path{arXiv:2102.10076}},
  \href {https://doi.org/10.1007/JHEP06(2021)145}
  {\path{doi:10.1007/JHEP06(2021)145}}.

\bibitem{CMS:2019eih}
A.~M. Sirunyan, et~al., {Measurement of the
  $\mathrm{t\bar{t}}\mathrm{b\bar{b}}$ production cross section in the all-jet
  final state in pp collisions at $\sqrt{s} =$ 13 TeV}, Phys. Lett. B 803
  (2020) 135285 (2020).
\newblock \href {http://arxiv.org/abs/1909.05306} {\path{arXiv:1909.05306}},
  \href {https://doi.org/10.1016/j.physletb.2020.135285}
  {\path{doi:10.1016/j.physletb.2020.135285}}.

\bibitem{CMS:2020osd}
A.~M. Sirunyan, et~al., {Search for a light charged Higgs boson in the H$^\pm$
  $\to $ cs channel in proton-proton collisions at $\sqrt{s} =$ 13 TeV}, Phys.
  Rev. D 102~(7) (2020) 072001 (2020).
\newblock \href {http://arxiv.org/abs/2005.08900} {\path{arXiv:2005.08900}},
  \href {https://doi.org/10.1103/PhysRevD.102.072001}
  {\path{doi:10.1103/PhysRevD.102.072001}}.

\bibitem{CMS:2019rlz}
A.~M. Sirunyan, et~al., {Search for a charged Higgs boson decaying into top and
  bottom quarks in events with electrons or muons in proton-proton collisions
  at $ \sqrt{\mathrm{s}} $ = 13 TeV}, JHEP 01 (2020) 096 (2020).
\newblock \href {http://arxiv.org/abs/1908.09206} {\path{arXiv:1908.09206}},
  \href {https://doi.org/10.1007/JHEP01(2020)096}
  {\path{doi:10.1007/JHEP01(2020)096}}.

\bibitem{CMS:2020imj}
A.~M. Sirunyan, et~al., {Search for charged Higgs bosons decaying into a top
  and a bottom quark in the all-jet final state of pp collisions at $ \sqrt{s}
  $ = 13 TeV}, JHEP 07 (2020) 126 (2020).
\newblock \href {http://arxiv.org/abs/2001.07763} {\path{arXiv:2001.07763}},
  \href {https://doi.org/10.1007/JHEP07(2020)126}
  {\path{doi:10.1007/JHEP07(2020)126}}.

\bibitem{Borodulin:2017pwh}
V.~I. Borodulin, R.~N. Rogalyov, S.~R. Slabospitskii, {CORE 3.1 (COmpendium of
  RElations, Version 3.1)} (2 2017).
\newblock \href {http://arxiv.org/abs/1702.08246} {\path{arXiv:1702.08246}}.

\bibitem{Slabospitsky:2002ag}
S.~R. Slabospitsky, L.~Sonnenschein, {TopReX generator (version 3.25): Short
  manual}, Comput. Phys. Commun. 148 (2002) 87--102 (2002).
\newblock \href {http://arxiv.org/abs/hep-ph/0201292}
  {\path{arXiv:hep-ph/0201292}}, \href
  {https://doi.org/10.1016/S0010-4655(02)00471-X}
  {\path{doi:10.1016/S0010-4655(02)00471-X}}.

\bibitem{CMS:2017wtu}
A.~M. Sirunyan, et~al., {Identification of heavy-flavour jets with the CMS
  detector in pp collisions at 13 TeV}, JINST 13~(05) (2018) P05011 (2018).
\newblock \href {http://arxiv.org/abs/1712.07158} {\path{arXiv:1712.07158}},
  \href {https://doi.org/10.1088/1748-0221/13/05/P05011}
  {\path{doi:10.1088/1748-0221/13/05/P05011}}.

\bibitem{Ma:1997up}
E.~Ma, D.~P. Roy, J.~Wudka, {Enhanced three-body decay of the charged Higgs
  boson}, Phys. Rev. Lett. 80 (1998) 1162--1165 (1998).
\newblock \href {http://arxiv.org/abs/hep-ph/9710447}
  {\path{arXiv:hep-ph/9710447}}, \href
  {https://doi.org/10.1103/PhysRevLett.80.1162}
  {\path{doi:10.1103/PhysRevLett.80.1162}}.

\bibitem{Decker:1992wz}
R.~Decker, M.~Nowakowski, A.~Pilaftsis, {Dominant three-body decays of a heavy
  Higgs and top quark}, Z. Phys. C 57 (1993) 339--348 (1993).
\newblock \href {http://arxiv.org/abs/hep-ph/9301283}
  {\path{arXiv:hep-ph/9301283}}, \href {https://doi.org/10.1007/BF01565067}
  {\path{doi:10.1007/BF01565067}}.

\bibitem{Mahlon:1994us}
G.~Mahlon, S.~J. Parke, {Finite width effects in top quark decays}, Phys. Lett.
  B 347 (1995) 394--398 (1995).
\newblock \href {http://arxiv.org/abs/hep-ph/9412250}
  {\path{arXiv:hep-ph/9412250}}, \href
  {https://doi.org/10.1016/0370-2693(95)00083-W}
  {\path{doi:10.1016/0370-2693(95)00083-W}}.

\bibitem{Jenkins:1996zd}
E.~E. Jenkins, {The Rare top decays $t \to b W^{+} Z$ and $t \to c W^{+}
  W^{-}$}, Phys. Rev. D 56 (1997) 458--466 (1997).
\newblock \href {http://arxiv.org/abs/hep-ph/9612211}
  {\path{arXiv:hep-ph/9612211}}, \href
  {https://doi.org/10.1103/PhysRevD.56.458}
  {\path{doi:10.1103/PhysRevD.56.458}}.

\bibitem{Altarelli:2000nt}
G.~Altarelli, L.~Conti, V.~Lubicz, {The t ---\ensuremath{>} WZ b decay in the
  standard model: A Critical reanalysis}, Phys. Lett. B 502 (2001) 125--132
  (2001).
\newblock \href {http://arxiv.org/abs/hep-ph/0010090}
  {\path{arXiv:hep-ph/0010090}}, \href
  {https://doi.org/10.1016/S0370-2693(00)01333-2}
  {\path{doi:10.1016/S0370-2693(00)01333-2}}.

\bibitem{Papaefstathiou:2017xuv}
A.~Papaefstathiou, G.~Tetlalmatzi-Xolocotzi, {Rare top quark decays at a 100
  TeV proton\textendash{}proton collider: $t \rightarrow bWZ$ and $t\rightarrow
  hc$}, Eur. Phys. J. C 78~(3) (2018) 214 (2018).
\newblock \href {http://arxiv.org/abs/1712.06332} {\path{arXiv:1712.06332}},
  \href {https://doi.org/10.1140/epjc/s10052-018-5701-8}
  {\path{doi:10.1140/epjc/s10052-018-5701-8}}.

\bibitem{Onyisi:2017mza}
P.~Onyisi, A.~Webb, {Impact of rare decays $t \to \ell' \nu b \ell\ell$ and $t
  \to q q' b \ell\ell$ on searches for top-associated physics}, JHEP 02 (2018)
  156 (2018).
\newblock \href {http://arxiv.org/abs/1704.07343} {\path{arXiv:1704.07343}},
  \href {https://doi.org/10.1007/JHEP02(2018)156}
  {\path{doi:10.1007/JHEP02(2018)156}}.

\bibitem{Quintero:2014lqa}
N.~Quintero, J.~L. Diaz-Cruz, G.~Lopez~Castro, {Lepton pair emission in the top
  quark decay $t \to bW^+\ell^-\ell^+$}, Phys. Rev. D 89~(9) (2014) 093014
  (2014).
\newblock \href {http://arxiv.org/abs/1403.3044} {\path{arXiv:1403.3044}},
  \href {https://doi.org/10.1103/PhysRevD.89.093014}
  {\path{doi:10.1103/PhysRevD.89.093014}}.

\bibitem{Bredenstein:2008zb}
A.~Bredenstein, A.~Denner, S.~Dittmaier, S.~Pozzorini, {NLO QCD corrections to
  t anti-t b anti-b production at the LHC: 1. Quark-antiquark annihilation},
  JHEP 08 (2008) 108 (2008).
\newblock \href {http://arxiv.org/abs/0807.1248} {\path{arXiv:0807.1248}},
  \href {https://doi.org/10.1088/1126-6708/2008/08/108}
  {\path{doi:10.1088/1126-6708/2008/08/108}}.

\bibitem{Cascioli:2013era}
F.~Cascioli, P.~Maierh\"ofer, N.~Moretti, S.~Pozzorini, F.~Siegert, {NLO
  matching for $t\bar t b \bar b$ production with massive $b$-quarks}, Phys.
  Lett. B 734 (2014) 210--214 (2014).
\newblock \href {http://arxiv.org/abs/1309.5912} {\path{arXiv:1309.5912}},
  \href {https://doi.org/10.1016/j.physletb.2014.05.040}
  {\path{doi:10.1016/j.physletb.2014.05.040}}.

\bibitem{Garzelli:2014aba}
M.~V. Garzelli, A.~Kardos, Z.~Tr\'ocs\'anyi, {Hadroproduction of
  $t\bar{t}b\bar{b}$ final states at LHC: predictions at NLO accuracy matched
  with Parton Shower}, JHEP 03 (2015) 083 (2015).
\newblock \href {http://arxiv.org/abs/1408.0266} {\path{arXiv:1408.0266}},
  \href {https://doi.org/10.1007/JHEP03(2015)083}
  {\path{doi:10.1007/JHEP03(2015)083}}.

\bibitem{Bevilacqua:2011aa}
G.~Bevilacqua, M.~Czakon, C.~G. Papadopoulos, M.~Worek, {Hadronic top-quark
  pair production in association with two jets at Next-to-Leading Order QCD},
  Phys. Rev. D 84 (2011) 114017 (2011).
\newblock \href {http://arxiv.org/abs/1108.2851} {\path{arXiv:1108.2851}},
  \href {https://doi.org/10.1103/PhysRevD.84.114017}
  {\path{doi:10.1103/PhysRevD.84.114017}}.

\bibitem{Bevilacqua:2014qfa}
G.~Bevilacqua, M.~Worek, {On the ratio of $ t\overline{t} b\overline{b} $ and $
  t\overline{t} jj $ cross sections at the CERN Large Hadron Collider}, JHEP 07
  (2014) 135 (2014).
\newblock \href {http://arxiv.org/abs/1403.2046} {\path{arXiv:1403.2046}},
  \href {https://doi.org/10.1007/JHEP07(2014)135}
  {\path{doi:10.1007/JHEP07(2014)135}}.

\bibitem{Bevilacqua:2017cru}
G.~Bevilacqua, M.~V. Garzelli, A.~Kardos, {$t\bar{t}b\bar{b}$ hadroproduction
  with massive bottom quarks with PowHel} (9 2017).
\newblock \href {http://arxiv.org/abs/1709.06915} {\path{arXiv:1709.06915}}.

\bibitem{Jezo:2018yaf}
T.~Je\v{z}o, J.~M. Lindert, N.~Moretti, S.~Pozzorini, {New NLOPS predictions
  for $t \bar{t} +b$ -jet production at the LHC}, Eur. Phys. J. C 78~(6) (2018)
  502 (2018).
\newblock \href {http://arxiv.org/abs/1802.00426} {\path{arXiv:1802.00426}},
  \href {https://doi.org/10.1140/epjc/s10052-018-5956-0}
  {\path{doi:10.1140/epjc/s10052-018-5956-0}}.

\bibitem{CMS:2020grm}
A.~M. Sirunyan, et~al., {Measurement of the cross section for
  $\text{t}\bar{\text{t}}$ production with additional jets and b jets in pp
  collisions at $\sqrt{s}=$ 13 TeV}, JHEP 07 (2020) 125 (2020).
\newblock \href {http://arxiv.org/abs/2003.06467} {\path{arXiv:2003.06467}},
  \href {https://doi.org/10.1007/JHEP07(2020)125}
  {\path{doi:10.1007/JHEP07(2020)125}}.

\bibitem{Grancagnolo:2019uvn}
S.~Grancagnolo, {Top quark pair-production cross section measurements at LHC},
  PoS DIS2019 (2019) 152 (2019).
\newblock \href {https://doi.org/10.22323/1.352.0152}
  {\path{doi:10.22323/1.352.0152}}.

\end{thebibliography}
